\documentclass[journal]{IEEEtran}

\usepackage{inputenc}
\usepackage{mathtools}
\usepackage{color}
\usepackage{array}
\usepackage{verbatim}
\usepackage{float}
\usepackage{amsmath}
\usepackage{amsthm}
\usepackage{amssymb}
\usepackage{graphicx}
\usepackage{longtable}
\usepackage{multirow}
\usepackage{balance}
\usepackage{mathtools}
\usepackage{tikz}
\usepackage{pgf-pie}

\usepackage[unicode=true,
bookmarks=false,
breaklinks=false,pdfborder={0 0 1},colorlinks=false]
{hyperref}
\hypersetup{
	colorlinks,bookmarksopen,bookmarksnumbered,citecolor=blue,urlcolor=blue}
\usepackage{cite}
\usepackage{nomencl} 
\makenomenclature
\usepackage{longtable}

\usepackage{indentfirst}
\usepackage{textcomp}
\usepackage[T1]{fontenc}
\usepackage{amsthm}
\usepackage{amsmath,amstext,amsfonts}
\usepackage{amssymb}
\usepackage{mathtools}
\usepackage[cmintegrals]{newtxmath}
\usepackage{adjustbox}
\usepackage{caption}
\usepackage{bm}  
\usepackage{siunitx}
\usepackage{physics}

\usepackage{graphicx}
\graphicspath{%
	{Figures/}
	{Figures/TiKz/}
	{Figures/Ipe/}
	{Figures/XFig/}
	{Figures/Inkscape/}
	{Figures/Matlab/}
	{Figures/Scanned/}
	{Figures/Saved/}
}
\usepackage{tikz}
\usepackage[list=true,labelformat=simple]{subcaption}
\usepackage{cite}
\usepackage{url}
\usepackage{xcolor}
\usepackage{siunitx}

\usepackage[inline]{enumitem}

\usepackage{algorithm,algpseudocode,float}
\usepackage{setspace} 

\algnewcommand{\algorithmicand}{\textbf{ and }}
\algnewcommand{\algorithmicor}{\textbf{ or }}
\algnewcommand{\AlgAnd}{\algorithmicand}
\algnewcommand{\AlgOr}{\algorithmicor}

\usepackage[noabbrev]{cleveref}  
\Crefname{figure}{Fig.}{Figs.}

\usepackage{tabularx,booktabs}
\newcolumntype{C}{>{\centering\arraybackslash}X} 
\usepackage{lipsum}

\makeatletter
\let\oldforeign@language\foreign@language
\DeclareRobustCommand{\foreign@language}[1]{%
	\lowercase{\oldforeign@language{#1}}}

\floatstyle{ruled}
\newfloat{algorithm}{tbp}{loa}
\providecommand{\algorithmname}{Algorithm}
\floatname{algorithm}{\protect\algorithmname}

\let\oldforeign@language\foreign@language
\DeclareRobustCommand{\foreign@language}[1]{%
	\lowercase{\oldforeign@language{#1}}}

\ifCLASSINFOpdf
\else
\fi

\newcommand{\AUCaeOnly}{0.561}      
\newcommand{\AUCdenoiseAE}{0.586}   
\newcommand{\AUCcondOnly}{0.827}    
\newcommand{\AUCfull}{0.884}        
\newcommand{\PRECaeOnly}{0.771}
\newcommand{\RECaeOnly}{0.274}
\newcommand{\FoneAeOnly}{0.404}
\newcommand{\PRECdenoiseAE}{0.764}
\newcommand{\RECdenoiseAE}{0.252}
\newcommand{\FoneDenoiseAE}{0.379}
\newcommand{\PREccondOnly}{0.863}
\newcommand{\RECcondOnly}{0.293}
\newcommand{\FoneCondOnly}{0.437}
\newcommand{\ParamsDenoise}{145.3~M}
\newcommand{\ParamsCCAE}{1.04~M}
\newcommand{\ParamsTotal}{146.4~M}
\newcommand{\FlopsDenoise}{69.9~G}
\newcommand{\FlopsCCAE}{0.98~G}
\newcommand{\FlopsTotal}{70.9~G}

\newtheorem{rem}{Remark}

\ifCLASSINFOpdf
\else
\fi

\def\ps@IEEEtitlepagestyle{%
	\def\@oddhead{\parbox[t][\height][t]{\textwidth}{\centering \scriptsize
			Personal use of this material is permitted. Permission from the author(s) and/or copyright holder(s), must be obtained for all other uses. Please contact us and provide details if you believe this document breaches copyrights.\\
			\noindent\makebox[\linewidth]{}
		}\hfil\hbox{}}%
	\def\@evenhead{\scriptsize\thepage \hfil \leftmark\mbox{}}%
	\def\@oddfoot{\parbox[t][\height][l]{\textwidth}{
			\vspace{-20pt}{\rule{\textwidth}{0.4pt}}\\ \footnotesize			{\bf{\footnotesize\textcolor{red}{A. M. Ali, Z. Tang, N. Nazarbay, and H. A. Hashim, "Unsupervised Maneuver-Aware Acoustic Fault Detection for Autonomous Drones," IEEE Transactions on Instrumentation and Measurement, 2026.}}} doi: \href{https://doi.org/10.1109/TIM.2026.3742099}{10.1109/TIM.2026.3742099}\\
			\noindent\makebox[\linewidth]
		}\hfil\hbox{}}%
	\def\@evenfoot{\MYfooter}}

\makeatother
\begin{document}
	\bstctlcite{IEEEexample:BSTcontrol}

\title{Unsupervised Maneuver-Aware Acoustic Fault Detection for Autonomous Drones}

\author{Ali M Ali, Ziyi Tang, Nurdaulet Nazarbay, and Hashim A. Hashim
\thanks{This work was supported in part by the National Sciences and Engineering
	Research Council of Canada (NSERC), under grant number RGPIN-2022-04937 and MITACS Globalink Research Internship Program Canada.}%
\thanks{A. M Ali and H. A. Hashim are with the Department of Mechanical and
	Aerospace Engineering, Carleton University, Ottawa, ON, K1S-5B6, Canada, Z. Tang is with Department of Electrical and Computer Engineering at the University of Alberta, Edmonton, AB, Canada, and N. Nazar is with The Hong Kong Polytechnic University, Hong Kong
	(e-mail: alimohamedali@cmail.carleton.ca, Ztang13@ualberta.ca, nurdaulet.nazarbay@connect.polyu.hk and hhashim@carleton.ca).}
\thanks{A. M. Ali and Z. Tang contributed equally to this work.}

}



\maketitle
\begin{abstract}
This paper presents a maneuver-aware acoustic fault detection framework for autonomous drones that integrates Noise2Noise-inspired deep learning denoising with maneuver-conditioned reconstruction. A key practical constraint motivating this work is that labeled faulty-flight data are difficult and potentially unsafe to collect; the proposed framework therefore follows an unsupervised learning paradigm in which only nominal flight recordings are required for training. In flight environments, acoustic signals acquired from unmanned aerial vehicles are subject to variability arising both from environmental noise and from structured, maneuver-dependent aerodynamic effects. To address these challenges simultaneously, a two-stage learning architecture is developed. In the first stage, a Noise2Noise-inspired denoising model attenuates stochastic acoustic noise while preserving fault-relevant spectral-temporal structures, without requiring clean reference signals. In the second stage, a maneuver-Conditioned Convolutional AutoEncoder (maneuver-CCAE) is trained using maneuver-related labels including drone type and flight direction to model nominal acoustic behavior under varying operating conditions. Fault detection is subsequently performed using reconstruction error as an anomaly score. Experimental results demonstrate that the proposed maneuver-aware conditioning raises the area under the ROC curve (AUC) from $\AUCaeOnly$ (unconditioned baseline) to $\AUCfull$ (full model), validating the critical role of maneuver-dependent modeling. The complete framework is deployed on an NVIDIA Jetson Orin Nano Super embedded platform within a ROS2 pipeline, achieving an end-to-end fault detection latency of approximately $20\,\text{ms}$ per audio segment with a TensorRT half-precision (FP16) backend, confirming real-time viability for onboard UAV health monitoring.
\end{abstract}


\section{Introduction}
\begin{figure}[h]
	\centering\includegraphics[width=9cm,height=6cm,keepaspectratio]{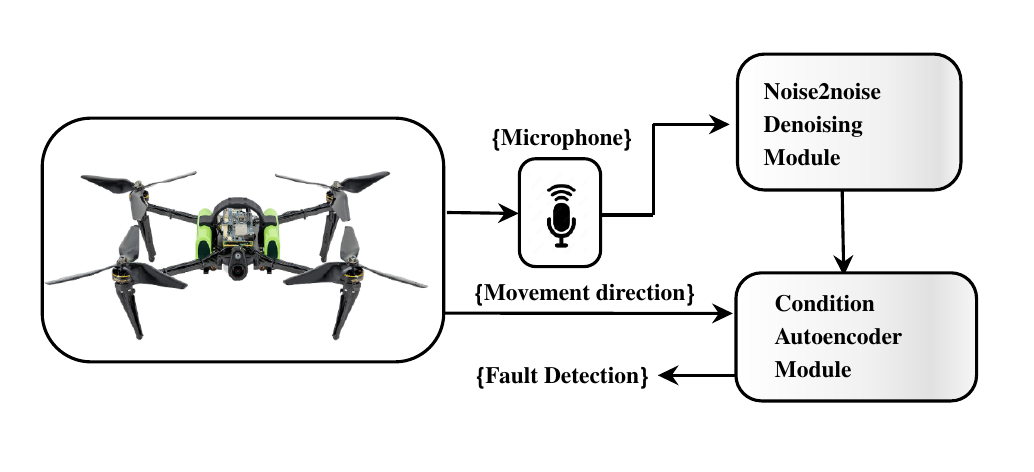}
	\caption{Overview of the proposed maneuver-aware acoustic fault detection framework. An onboard microphone feeds a Noise2Noise-inspired denoising module, whose output is processed by a maneuver-conditioned autoencoder; reconstruction error serves as the anomaly score.}
	\label{Figure_0}
\end{figure}

Autonomous drones in safety-critical roles such as infrastructure inspection, search and rescue, and logistics fail predominantly through propulsion faults such as propeller damage and motor degradation \cite{hash2025_RIENG_Avionics}. These faults reshape the acoustic field in structured ways, including attenuated dominant peaks, spectral spreading, and sideband generation, that a single onboard microphone can detect passively. The difficulty is that these same acoustic features are also modulated by normal flight maneuvers, so a healthy drone banking or yawing can look, acoustically, like a faulty one flying straight, which is the central problem this work addresses. In this work, a machine learning-based module is developed to detect and flag fault instances using only acoustic signals captured by a microphone mounted on the drone. The considered fault scenarios include single-motor degradation and propeller damage (e.g., broken or chipped propeller blades); see Fig.~\ref{Figure_0}.

A key practical constraint motivating this work is that labeled faulty-flight data are difficult and potentially unsafe to collect. The proposed framework therefore follows a strictly unsupervised learning paradigm: only nominal (healthy) flight recordings are required for training, and fault detection is performed entirely through reconstruction error at inference time, without access to any faulty examples during training or threshold selection. Despite its appeal, reliable acoustic Fault Detection Diagnosis (FDD) for drones operating in real flight environments remains challenging. Drone acoustic signals are inherently non-stationary and are strongly influenced by environmental noise, wind disturbances, and maneuver-dependent aerodynamic effects. The same healthy vehicle can therefore produce markedly different spectral-temporal signatures across flight directions, speeds, and operating conditions. Maneuvers alter aerodynamic loading, flow interaction, and propulsion dynamics, producing systematic changes in acoustic emissions even under healthy operation. As a result, most existing reconstruction-based frameworks implicitly treat maneuver-induced variability as nuisance noise rather than structured information. Consequently, acoustic variations induced by normal flight maneuvers may be misinterpreted as faults, leading to increased false-alarm rates and reduced operational trust. This paper addresses these limitations by proposing a maneuver-conditioned, reconstruction-based acoustic fault detection framework for autonomous drones that explicitly models maneuver-dependent variability while maintaining robustness to environmental noise.

Early acoustic fault detection approaches were largely based on deterministic signal processing, exploiting time-domain statistics, spectral features, and time-frequency representations to identify deviations from nominal behavior \cite{ref01,ref02,ref03,ref04,ref05}. These methods, while computationally efficient, often rely on assumptions of stationarity or repeatability that are violated during aggressive or sustained maneuvers. Recently, supervised learning techniques have become popular \cite{Sam2026_RoofBinary,hash2026_MFWAV}, approaching fault diagnosis as a classification problem using features created by humans or deep neural networks \cite{ref14,ref15,ref16,ref17}. Although these methods can achieve high accuracy in controlled settings, they require a large amount of labeled fault data, the collection of which is costly, risky, and often impractical for real-world drone operations. In contrast, unsupervised reconstruction-based methods offer a useful alternative by learning models of normal behavior from healthy data and detecting anomalies through reconstruction errors \cite{ref63,ref64,ref65,ref66}. Such approaches are well suited to real deployments because they do not require prior knowledge of fault types and can detect previously unseen abnormalities. However, as noted above, most existing frameworks fail to account for structured maneuver-induced variability, severely degrading detection performance in real-flight settings.

\paragraph*{Structure}
The rest of this paper is structured as follows. Section~\ref{sec:Related-Work} presents related work and contributions. Section~\ref{sec:problem_formulation} presents the problem formulation and acoustic feature representation. Section~\ref{sec: main} introduces the proposed two-stage maneuver-conditioned learning architecture. Section~\ref{sec: Results} presents the results and discussion. Finally, Section~\ref{sec: Conclusion} provides concluding remarks.

\section{Related Work\label{sec:Related-Work}}

\subsection{Deterministic Signal-Processing Based Acoustic FDD}

Prior to the widespread adoption of machine learning, acoustic FDD relied predominantly on deterministic, physics-based signal processing. These methods detect faults by monitoring changes in time-domain statistics (e.g., Root Mean Square (RMS) energy, kurtosis, crest factor), frequency-domain signatures (e.g., power spectral density), or joint time-frequency representations \cite{ref01,ref02,ref03,ref04}. Envelope analysis and cepstrum-based techniques are particularly effective for identifying periodic impulsive faults, such as bearing damage or gear mesh anomalies \cite{ref03,ref04}. To address speed variability, order tracking methods transform frequency spectra into speed-normalized representations \cite{ref05,ref06}. In Unmanned Aerial Vehicles (UAVs) applications, lightweight Fast Fourier Transform (FFT)-based monitoring and spectral thresholding have been employed for detecting propeller imbalance and motor faults \cite{ref08,ref09}. While deterministic methods offer interpretability and low computational cost, their performance is often sensitive to noise, recording conditions, and maneuver-induced aerodynamic variability.

\subsection{Supervised Learning for Acoustic Fault Diagnosis}

Supervised learning approaches formulate fault diagnosis as a classification problem using labeled examples of healthy and faulty states. Traditional machine learning pipelines typically rely on handcrafted features such as Mel Frequency Cepstral Coefficients (MFCC), spectral entropy, and time-frequency descriptors, combined with classifiers including Support Vector Machine (SVM), decision trees, $k$-Nearest Neighbors (KNN), and ensemble methods \cite{ref14,ref15,ref29,ref30}. Deep learning has significantly advanced supervised acoustic FDD by enabling end-to-end learning from raw waveforms or spectrogram representations \cite{ref16,ref33,ref34}. CNN-based models have demonstrated strong performance in detecting propeller damage and rotor impairments \cite{ref35}. Recurrent architectures including Long Short-Term Memory (LSTM), Gated Recurrent Unit (GRU), and Bidirectional LSTM (BiLSTM) networks effectively capture temporal dependencies \cite{ref26,ref27,ref37}. Despite these advances, supervised approaches face fundamental scalability limitations: acquiring labeled fault data typically requires artificially inducing damage, which is costly and potentially unsafe, and models may generalize poorly to new operating conditions.

\subsection{Unsupervised Anomaly Detection and Reconstruction-Based Methods}

Unsupervised anomaly detection mitigates the need for labeled fault data by identifying deviations from nominal behavior \cite{ref38}. Reconstruction-based anomaly detection has emerged as a dominant paradigm for acoustic monitoring. Autoencoders and their variants, including convolutional, variational, denoising, and masked autoencoders, learn compact representations of healthy signals and use reconstruction error as an anomaly score \cite{ref63,ref64,ref65,ref66,ref67}. Such methods have been successfully applied to UAV propeller monitoring \cite{ref63}, wind turbine diagnostics \cite{ref64,zhang2022multiclass}, and industrial machinery fault detection using benchmark datasets such as Malfunctioning Industrial Machine Investigation and Inspection (MIMII) \cite{ref65}. To improve generalization, recent studies have explored lightweight autoencoder architectures for embedded deployment \cite{ref67} and conditional or hierarchical models that incorporate auxiliary information into the reconstruction process \cite{ref66}. These results suggest that explicitly modeling structured sources of variability can significantly enhance anomaly detection performance.

\subsection{Maneuver-Dependent Variability: An Open Challenge}

A persistent limitation across much of the acoustic FDD literature is the inadequate treatment of maneuver-dependent variability in drone acoustics. Maneuvers alter aerodynamic loading, flow interaction, and propulsion dynamics, producing systematic changes in acoustic emissions even under healthy operation. When such effects are not explicitly modeled, they can be misinterpreted as anomalies, leading to false alarms and reduced operational reliability. While some supervised multi-class systems have incorporated maneuver information \cite{ref28}, maneuver conditioning remains largely unexplored in unsupervised reconstruction-based frameworks. This gap motivates the present work.

\subsection{Contributions}
Unlike deterministic signal-processing approaches \cite{ref12,ref10}, the proposed framework does not require fixed spectral signatures or threshold tuning and remains effective under highly non-stationary flight conditions. Unlike supervised methods \cite{ref14,ref37}, it operates in an unsupervised manner and requires only nominal flight data for training. Although recent works \cite{xu2025lightweight,peng2021wind} have demonstrated effective deep learning-based acoustic anomaly detection, most consider stationary or quasi-stationary operating conditions. UAV acoustic signals exhibit significant non-stationarity due to maneuver-dependent aerodynamic effects, motivating the need for maneuver-conditioned diagnostic models. The contributions of this paper are as follows:
\begin{enumerate}
	\item
	UAV acoustic fault detection is formulated as a maneuver-conditioned
	unsupervised learning task, in which nominal acoustic behavior is
	explicitly modeled as a function of maneuver-related operating conditions
	(e.g., drone type and flight direction). This enables a systematic separation
	of maneuver-induced variability from fault-induced acoustic deviations
	and is, to the best of the authors' knowledge, the first such formulation
	for UAV acoustic fault detection.
	
	\item 
	A two-stage architecture integrates Noise2Noise-inspired acoustic
	denoising with a conditional convolutional autoencoder. The denoising
	stage attenuates stochastic environmental disturbances using
	maneuver-consistent surrogate pairs, without requiring clean reference
	signals, while the conditional reconstruction stage captures
	maneuver-dependent nominal dynamics through Conditional Batch
	Normalization (CBN).
	
	\item The complete pipeline, training, threshold selection, and
	inference, uses only healthy-flight acoustic data. The anomaly threshold
	is fixed at the 95th percentile of the nominal validation
	reconstruction-error distribution, so no faulty sample or label is
	examined at any stage.
	
	\item 
	The framework is deployed and evaluated on an NVIDIA Jetson Orin Nano
	Super embedded platform within a ROS2 pipeline, with the two networks
	exported to ONNX and executed as TensorRT half-precision (FP16) engines
	on the GPU Tensor Cores. This allows larger-capacity networks to run at an end-to-end latency of
	$\approx20$~ms per audio segment, confirming practical real-time
	viability for onboard UAV health monitoring.
\end{enumerate}
\section{Problem Formulation\label{sec:problem_formulation}}

This section formulates the acoustic anomaly detection problem considered in this work for quadrotor drones. The objective is to detect abnormal mechanical conditions, such as motor degradation or propeller damage, using acoustic signals recorded by an onboard microphone during flight. A key practical constraint of the problem is that only healthy-flight data are available during training, since collecting labeled faulty data can be costly and unsafe. Therefore, the proposed framework follows an unsupervised learning approach, where the model learns the nominal acoustic behavior of the drone under different maneuvering conditions. During operation, deviations from the learned nominal acoustic model are quantified using an anomaly score derived from the reconstruction error. This anomaly score allows the system to differentiate abnormal acoustic signatures caused by mechanical faults from normal maneuver-induced acoustic variations.

\subsection{Acoustic Feature Representation}

\begin{figure*}[h]
	\centering\includegraphics[width=18cm,height=12cm,keepaspectratio]{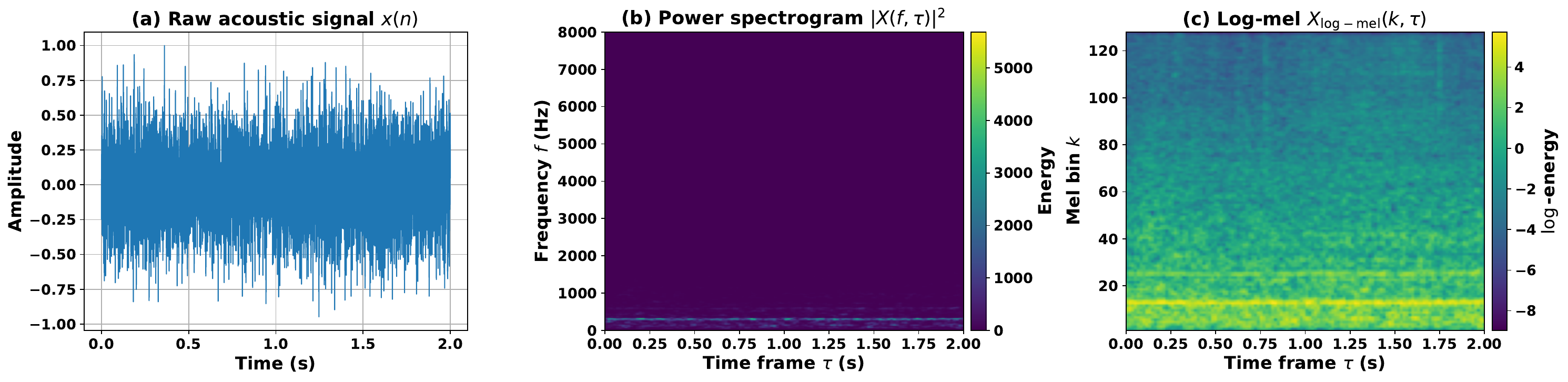}
	\caption{Feature extraction pipeline. $(a)$~Raw signal $x(n)$ at 16~kHz. $(b)$~Power spectrogram $|X(f,\tau)|^{2}$. $(c)$~Log-mel spectrogram $X_{\text{log-mel}}(k,\tau)$.}
	\label{Figure_1}
\end{figure*}

Each drone acoustic recording is represented using a log-mel spectrogram, which provides a compact and intuitive time-frequency characterization of the sound signal. The resulting discrete-time acoustic signal is defined as $x(n)$, $n=1,\ldots,N$, sampled at 16~kHz and segmented into a 2-second window, corresponding to $N=f_{s}\cdot T_{w}=32{,}000$ samples per segment, where $f_{s}=16{,}000$~Hz is the sampling frequency and $T_{w}=2$~s is the analysis window duration. This duration is sufficient to capture multiple propeller rotation cycles and maneuver-dependent acoustic patterns while maintaining adequate temporal resolution. The Short-Time Fourier Transform (STFT) of $x(n)$ is first computed using a sliding analysis window \cite{oppenheim1999discrete}:
\begin{equation}
	X(f,\tau)=\sum_{n}x(n)\,w(n-\tau)\,e^{-j2\pi fn},
\end{equation}
where $w(\cdot)$ denotes the analysis window, $f$ is the discrete frequency bin index, and $\tau$ is the time frame index. The power spectrogram $|X(f,\tau)|^{2}$ is mapped onto the mel scale using a bank of $F$ overlapping triangular filters \cite{smith2007mathematics}:
\begin{equation}
	S_{\text{mel}}(k,\tau)=\sum_{f}H_{k}(f)|X(f,\tau)|^{2},\quad k=1,\ldots,F,
	\label{eq:mel-spectrogram}
\end{equation}
where $H_{k}(f)$ denotes the $k$-th mel filter, and $k$ is the mel filter index. Logarithmic compression is then applied:
\begin{equation}
	X_{\text{log-mel}}(k,\tau)=\log(S_{\text{mel}}(k,\tau)),
	\label{eq:log-mel spectrum}
\end{equation}
yielding the matrix $X\in\mathbb{R}^{F\times T}$ used throughout the framework, where $t=1,\ldots,T$ indexes the time frames in the spectrogram matrix. Fig.~\ref{Figure_1} illustrates the transformation from raw waveform to log-mel spectrogram.

\subsection{Unsupervised Acoustic Anomaly Detection Objective}

During training, only samples corresponding to nominal (healthy) drone operation are available:
\begin{equation}
	\mathcal{D}_{\text{train}}=\{X_{i}\}_{i=1}^{N_{\text{tr}}},\quad X_{i}\in\mathbb{R}^{F\times T}.
\end{equation}
The model parameters $\theta$ are optimized by minimizing a reconstruction loss over nominal data:
\begin{equation}
	\min_{\theta}\frac{1}{N_{\text{tr}}}\sum_{i=1}^{N_{\text{tr}}}
	\mathcal{L}(X_{i},f_{\theta}(X_{i})),
	\label{eq:obective function without c}
\end{equation}
where $\mathcal{L}(\cdot,\cdot)$ is the reconstruction loss defined in Eq.~\eqref{eq:objective}. At inference time, the anomaly score is:
\begin{equation}
	\mathcal{A}(X^{*})=\mathcal{L}(X^{*},f_{\theta}(X^{*})),
	\label{eq:aonmly score without c}
\end{equation}
instantiated as the hybrid Mean Squared Error (MSE) MSE+SSIM loss. While effective for simple scenarios, this formulation implicitly assumes that all nominal acoustic variability is homogeneous. In practice, healthy drone acoustics exhibit strong, structured variations driven by maneuvering direction and operating conditions.

The choice of reconstruction loss as the anomaly scoring function is motivated by the strictly unsupervised constraint of this work. In the absence of labeled faulty data during training, no class-discriminative loss can be optimized directly. Reconstruction loss provides a principled self-supervised surrogate: the model is trained to compress and reconstruct nominal signals accurately, so that any input deviating from the learned nominal manifold will incur a higher reconstruction error. A higher reconstruction error indicates a greater deviation from the learned nominal acoustic manifold and is therefore more likely to correspond to a fault condition. This property is particularly well suited to log-mel spectrogram representations, where fault-induced deviations manifest as localized changes in spectral energy distribution. Alternative loss functions such as contrastive or metric learning losses require negative (faulty) samples during training and are therefore incompatible with the unsupervised constraint. Adversarial losses introduce training instability that is undesirable in resource-constrained embedded deployment. The hybrid MSE+SSIM formulation adopted here (see Eq.~\eqref{eq:objective}) penalizes both pointwise spectral differences and structural dissimilarity, preserving harmonic patterns important for acoustic fault characterization.

\subsection{Maneuver-Conditioned Acoustic Anomaly Detection}

Reconstruction-based unsupervised models implicitly assume that all nominal acoustic observations are generated from a single homogeneous distribution. This assumption is violated in drone platforms, where maneuvering actions induce systematic, repeatable changes in acoustic emissions even under healthy operation. To address this, the anomaly detection problem is reformulated as a maneuver-conditioned unsupervised learning task. Each acoustic observation is represented by the tuple $(X_{i},c_{i})$, where $c_{i}\in\mathcal{C}$ encodes structured operating information available during training, such as flight direction (e.g., forward, backward, left, right, clockwise, counterclockwise) or drone identity. The reconstruction model is extended to explicitly account for maneuvering conditions by defining a conditional mapping $f_{\theta}:\mathbb{R}^{F\times T}\times\mathcal{C}\rightarrow\mathbb{R}^{F\times T}$, where the output represents the reconstruction of $X_{i}$ given both the acoustic observation and its associated maneuver condition. The maneuver-conditioned training dataset is:
\begin{equation}
	\mathcal{D}_{\text{train}}^{\text{cond}}=\{(X_{i},c_{i})\}_{i=1}^{N_{\text{tr}}},
\end{equation}
where all samples correspond to nominal operation. The model parameters $\theta$ are learned by minimizing:
\begin{equation}
	\underset{\theta}{\min}\frac{1}{N_{\text{tr}}}\sum_{i=1}^{N_{\text{tr}}}
	\mathcal{L}(X_{i},f_{\theta}(X_{i},c_{i})).
\end{equation}
At inference time, given $\left(X^{*},c^{*}\right)$, the maneuver-conditioned anomaly score is:
\begin{equation}
	\mathcal{A}_{\text{cond}}\left(X^{*},c^{*}\right)=
	\mathcal{L}(X^{*},f_{\theta}(X^{*},c^{*})).
\end{equation}
Conditioning on the maneuver state reduces the sensitivity of the anomaly score to benign operating-condition variability, ensuring that elevated reconstruction errors are primarily associated with fault-related acoustic deviations.
\begin{figure}[h]
	\centering\includegraphics[width=9cm,height=6cm,keepaspectratio]{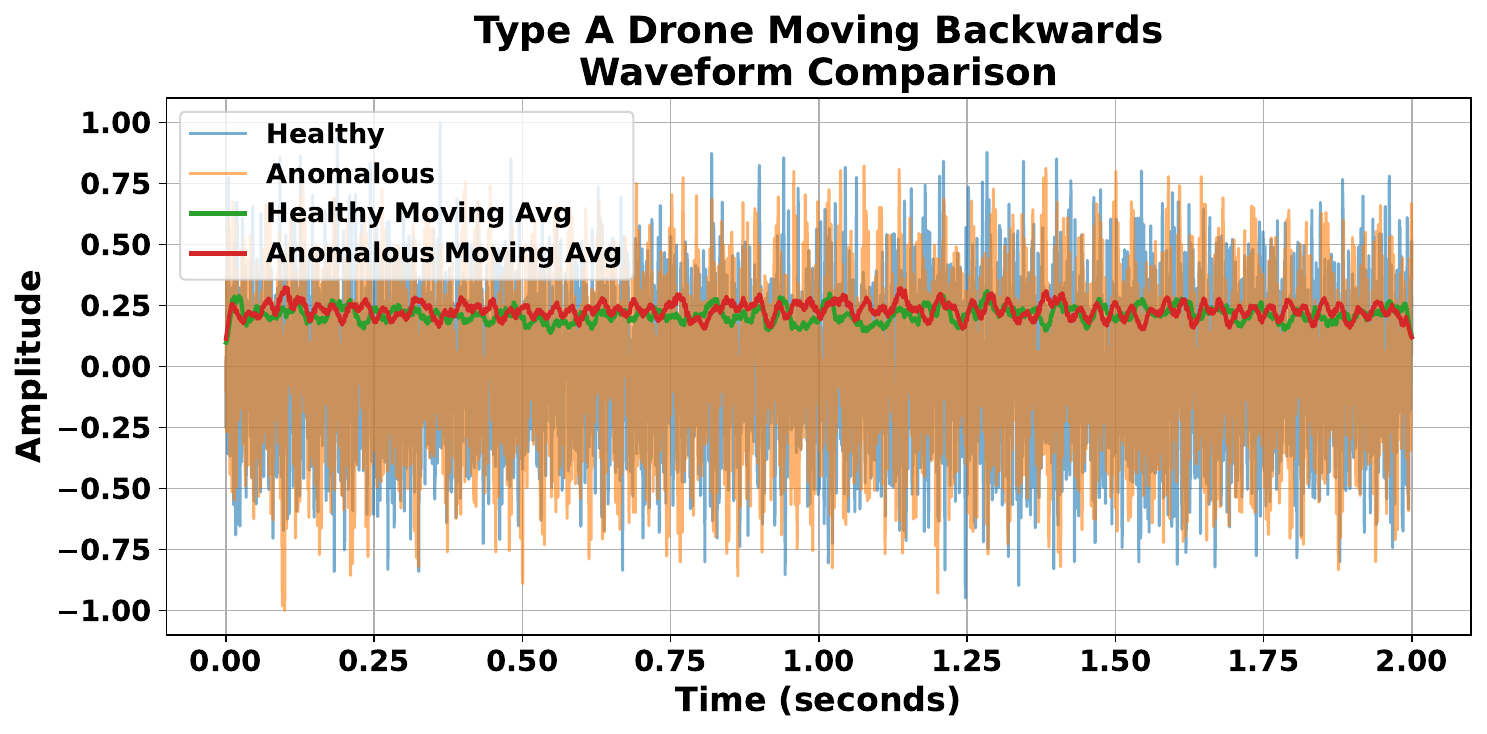}
	\caption{Time-domain comparison of healthy and anomalous acoustic signals with their moving-average envelopes.}
	\label{fig:wave}
\end{figure}

\begin{figure*}[h]
	\subfloat{\includegraphics[width=8.5cm,height=8.5cm,keepaspectratio]{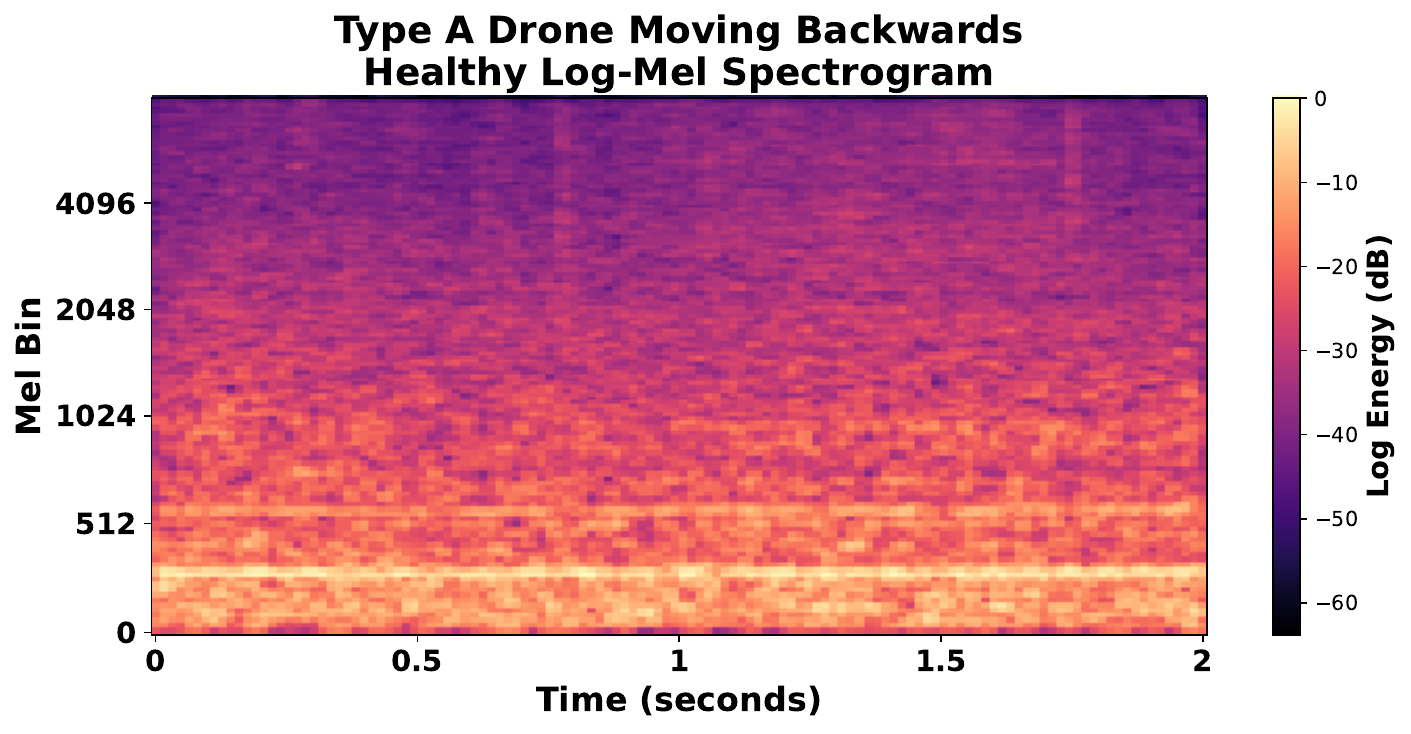}}
	\hfill{}
	\subfloat{\includegraphics[width=8.5cm,height=8.5cm,keepaspectratio]{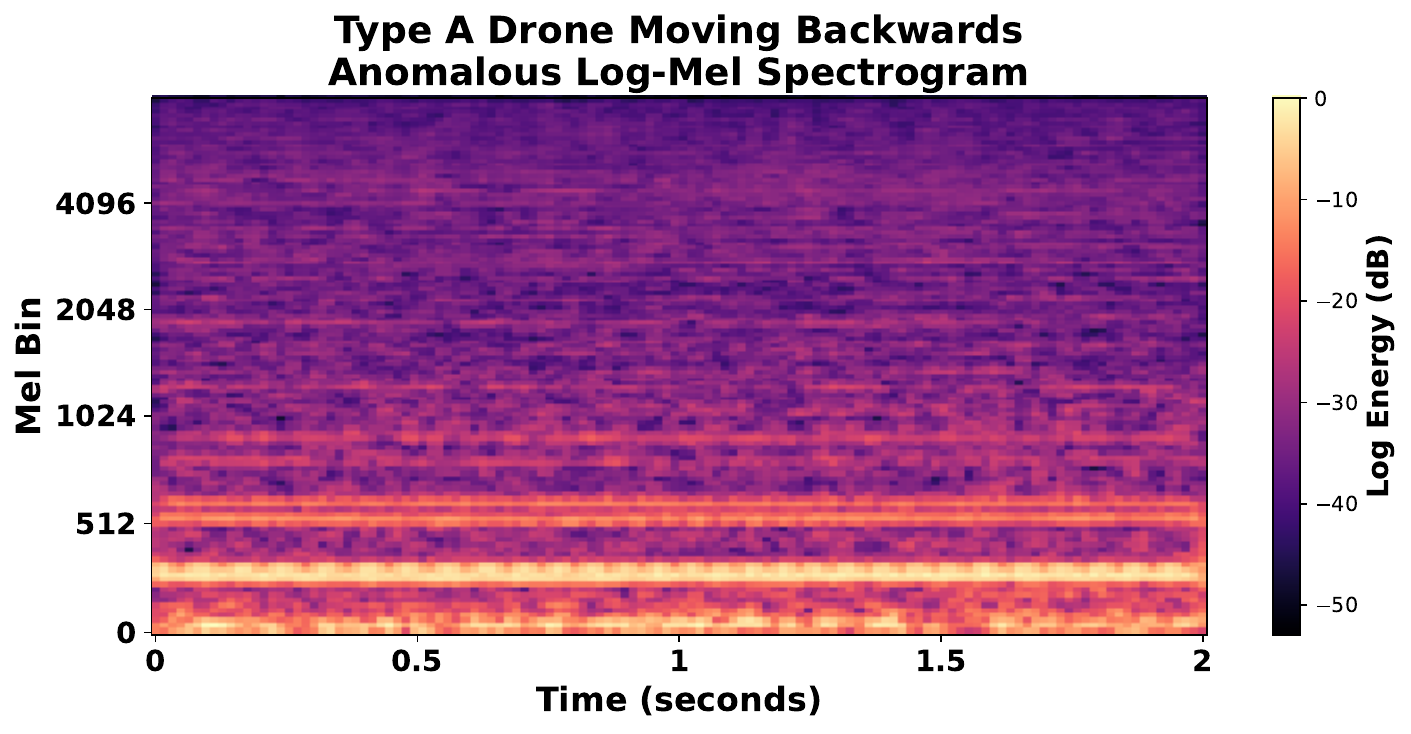}}
	\hfill{}
	\subfloat{\includegraphics[width=8.5cm,height=8.5cm,keepaspectratio]{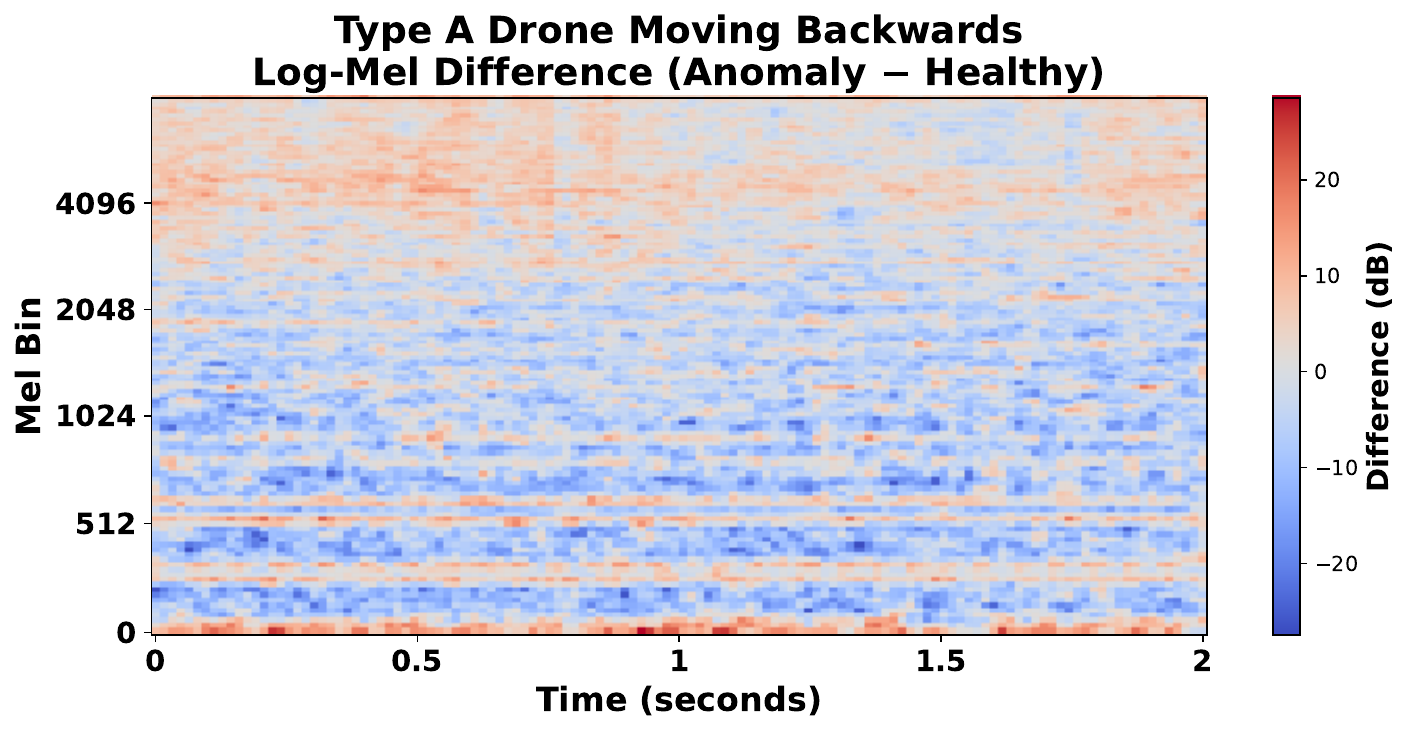}}
	\hfill{}
	\subfloat{\includegraphics[width=8.5cm,height=8.5cm,keepaspectratio]{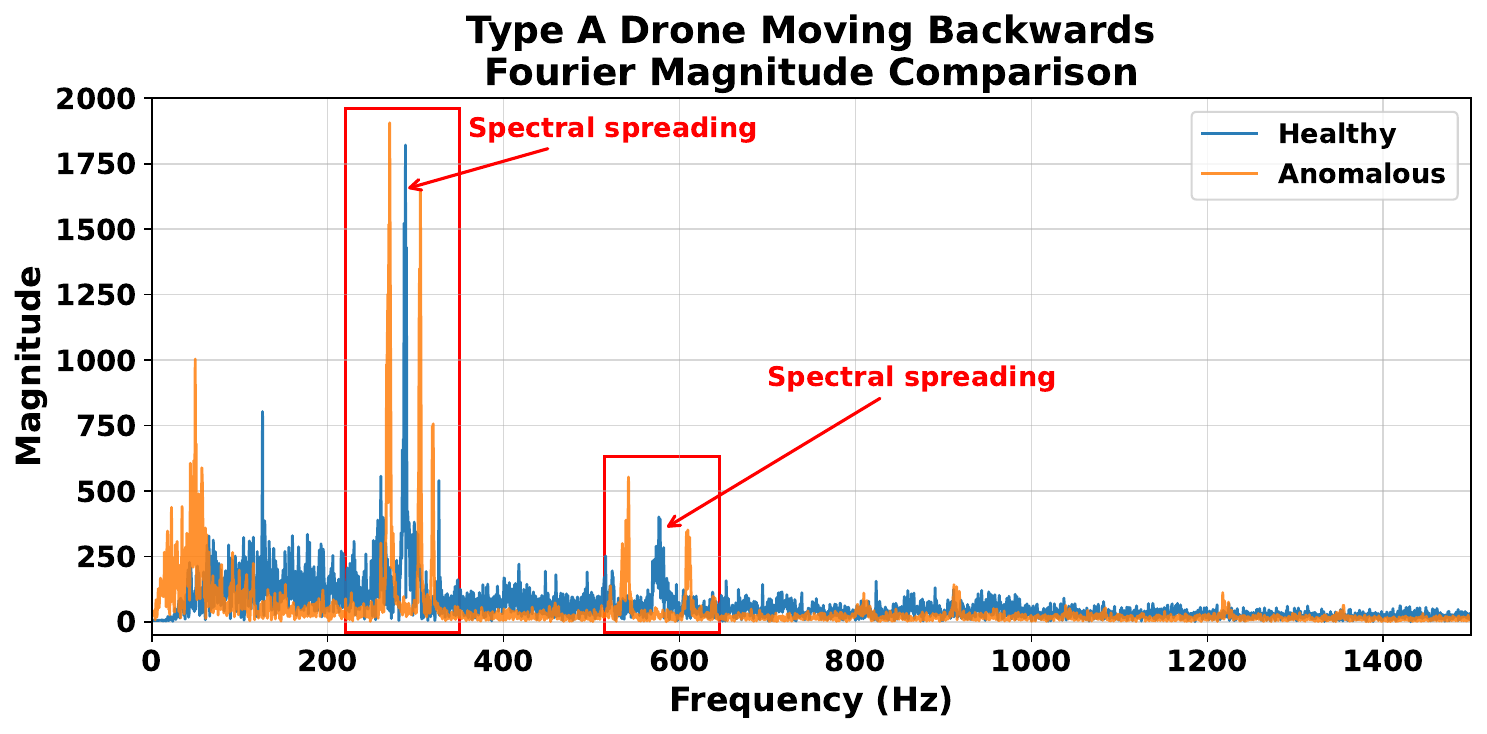}}
	\caption{Acoustic analysis of a backward maneuver (Type-A drone). $(a)$~Healthy log-mel spectrogram. $(b)$~Anomalous log-mel spectrogram. $(c)$~Difference map. $(d)$~Fourier magnitude spectra: the anomalous condition shows spectral spreading and additional sideband components caused by propulsion system faults.}
	\label{fig:figure_3}
\end{figure*}

\subsection{Training Data}

The dataset employed in this study is adopted from drone-sound data of Yi \textit{et al.}~\cite{yi2023sound}. It consists of acoustic recordings collected from three commercially available quadrotor platforms: the Holy Stone HS720 (Type~A), MJX Bugs~12~EIS (Type~B), and ZLRC SG906 Pro2 (Type~C). The sound signals were recorded using a R{\O}DE Wireless Go2 microphone mounted on top of the drone body in an anechoic chamber to eliminate wall reflections, and were originally sampled at 48~kHz and downsampled to 16~kHz. To capture the variability associated with different operating conditions, the drones were operated under six motion directions: forward, backward, left, right, clockwise, and counterclockwise. Fault conditions were generated by introducing controlled defects in the propulsion system, including partial cuts in the propeller blades and mechanical deformation of the motor cap, which induce abnormal vibration and altered acoustic patterns. To simulate real-flight conditions, the drone recordings were mixed with background noise at a signal-to-noise ratio (SNR) between $-5$ and $+5$~dB, using outdoor recordings and publicly available ambient noise sources. The waveform comparison in Fig.~\ref{fig:wave} and the spectral analysis in Fig.~\ref{fig:figure_3} illustrate why time-domain inspection alone is insufficient and why spectrogram representations are required for reliable fault detection. In particular, the healthy signal exhibits stronger dominant peaks corresponding to the nominal propulsion system frequencies. In contrast, the anomalous condition results in a redistribution of spectral energy, where the dominant peak magnitude decreases while additional frequency components appear around the main frequencies.

\section{Proposed Architecture\label{sec: main}}

This section presents the proposed two-stage learning architecture for maneuver-aware acoustic fault detection in autonomous drones. The framework addresses two fundamental challenges: (i)~strong stochastic disturbances arising from environmental noise and wind effects, and (ii)~structured, maneuver-dependent variability in nominal acoustic emissions. First, a Noise2Noise-inspired denoising network suppresses stochastic acoustic corruption without requiring clean reference signals. Second, a maneuver-conditioned convolutional autoencoder models nominal acoustic behavior and detects faults through reconstruction error.

\subsection{Noise2Noise-Inspired Denoising Network\label{subsec:Noise2Noise-based-denoising-netw}}

Since clean reference recordings cannot be collected in practice, the proposed framework employs a denoising model to suppress stochastic corruption while preserving fault-relevant spectral-temporal structure. The Noise2Noise framework, introduced by \cite{lehtinen2018noise2noise}, demonstrated that a denoising model can be effectively trained without clean ground-truth signals by learning from pairs of independently corrupted observations. Given two independent noise realizations:
\begin{alignat*}{1}
	x_{1} & =s+n_{1}, \\
	x_{2} & =s+n_{2},
\end{alignat*}
where $s\in\mathbb{R}^{n}$ is the unknown clean signal and $n_{1}$, $n_{2}$ are independent zero-mean noise realizations. Minimizing $\|f_{\theta}(x_{1})-x_{2}\|^{2}$ drives the model to recover the clean signal $s$. For our training dataset, let $\tilde{X}_{i}\in\mathbb{R}^{1\times F\times T}$ denote the noisy log-mel spectrogram of the $i$-th acoustic segment. Maneuver information is encoded using one-hot vectors $c_{i}^{\text{type}}\in\mathbb{R}^{3}$ (three drone platforms) and $c_{i}^{\text{dir}}\in\mathbb{R}^{6}$ (six maneuver directions) and projected into a $d_{e}$-dimensional latent embedding:
\begin{align}
	z_{i}^{\text{type}} & =W_{t}c_{i}^{\text{type}}+b_{t}, \\
	z_{i}^{\text{dir}} & =W_{d}c_{i}^{\text{dir}}+b_{d}.
\end{align}
With $d_{e}=8$, the concatenated 16-dimensional maneuver embedding $z_{i}$ conditions the denoising network via CBN:
\begin{align}
	h^{(l)} & = \sigma\!\left(\text{CBN}\!\left(W^{l}\ast h^{(l-1)}\right)\right), \\
	y_{b,c,h,w} & = \gamma_{c}(z_{i})\hat{h}_{b,c,h,w}+\beta_{c}(z_{i}),
\end{align}
where
\begin{align}
	\gamma(z_{i}) & =W_{\gamma}z_{i}+b_{\gamma}, \\
	\beta(z_{i}) & =W_{\beta}z_{i}+b_{\beta},
\end{align}
are maneuver-dependent affine parameters with $W_{\gamma},W_{\beta}\in\mathbb{R}^{C_{l}\times2d_{e}}$, where $C_{l}$ is the number of channels of layer $l$ and $2d_{e}$ is the dimension of the concatenated embedding $z_{i}=[z_{i}^{\text{type}};z_{i}^{\text{dir}}]$. The network architecture is summarized in Table~\ref{table3-1}.

Let $\bar{X}_{i}\in\mathbb{R}^{1\times F\times T}$ denote the output of the denoising network and let $X_{i}^{\text{ref}}\in\mathbb{R}^{1\times F\times T}$ be a maneuver-consistent surrogate target drawn from the training set. Throughout the paper, $\bar{X}$ denotes a denoised spectrogram (Stage~1 output) and $\hat{X}$ denotes a reconstruction produced by the autoencoder (Stage~2 output). The hybrid denoising loss is:
\begin{equation}
	\mathcal{L}=\alpha\,\text{MSE}(\bar{X}_{i},X_{i}^{\text{ref}})+(1-\alpha)\!\left(1-\text{SSIM}(\bar{X}_{i},X_{i}^{\text{ref}})\right),
	\label{eq:objective}
\end{equation}
with $\alpha=0.5$, where:
\begin{equation}
	\text{MSE}(\bar{X}_{i},X_{i}^{\text{ref}})=\frac{1}{FT}
	\sum_{f=1}^{F}\sum_{t=1}^{T}\!\left(\bar{X}_{i}(f,t)-X_{i}^{\text{ref}}(f,t)\right)^{2},
	\label{eq:MSE}
\end{equation}
\begin{equation}
	\text{SSIM}(\bar{X}_{i},X_{i}^{\text{ref}})=
	\frac{(2\mu_{\bar{X}_{i}}\mu_{X_{i}^{\text{ref}}}+C_{1})
		(2\sigma_{\bar{X}_{i}X_{i}^{\text{ref}}}+C_{2})}
	{(\mu_{\bar{X}_{i}}^{2}+\mu_{X_{i}^{\text{ref}}}^{2}+C_{1})
		(\sigma_{\bar{X}_{i}}^{2}+\sigma_{X_{i}^{\text{ref}}}^{2}+C_{2})}.
	\label{eq:SSIM}
\end{equation}
Here, $C_{1}=(0.01L)^{2}$ and $C_{2}=(0.03L)^{2}$ are stability constants with $L$ being the dynamic range of the spectrogram values. The SSIM term encourages preservation of local structural patterns and harmonic relationships that are important for acoustic fault characterization, while the MSE term captures pointwise spectral energy redistribution caused by fault-induced vibration modulation.

\begin{rem}
	Unlike the original Noise2Noise formulation, which assumes two independently corrupted realizations of the same underlying clean signal, the proposed implementation employs maneuver-consistent surrogate targets constructed from samples belonging to the same operating condition. As a result, the independence assumption of the original Noise2Noise theory is not strictly satisfied. Since true independent noise realizations of the same underlying signal are not available for in-flight drone recordings, maneuver-consistent surrogate pairs are used as training targets, following the spirit of the Noise2Noise paradigm under the assumption that the clean spectral-temporal structure remains approximately stable within the same maneuver condition. The proposed denoising stage should therefore be interpreted as a weak-reference Noise2Noise-inspired learning strategy that leverages structured similarity within maneuver classes. The objective is not to perfectly recover a single clean signal realization, but rather to learn a maneuver-consistent representation that suppresses stochastic environmental disturbances while preserving fault-relevant spectral-temporal structures.
\end{rem}

\begin{table*}[h]
	\caption{\label{table3-1}Denoising network architecture summary. Channel counts are given as input${\to}$output for each block; the feature-map size is halved at every encoder stage and doubled at every decoder stage, with skip connections concatenated from the corresponding encoder stage.}
	\centering{}{\footnotesize{}}%
	\begin{tabular}{>{\centering}p{1.1cm}>{\centering}p{1cm}>{\centering}m{1.5cm}>{\centering}m{1.5cm}>{\centering}m{1.4cm}>{\centering}m{1.8cm}>{\centering}m{1.2cm}>{\centering}m{1.5cm}>{\centering}m{1.5cm}>{\centering}m{1.2cm}}
		\toprule
		{\scriptsize{}Stage} & {\scriptsize{}Input} & {\scriptsize{}Encoder 1} & {\scriptsize{}Encoder 2} & {\scriptsize{}Encoder 3} & {\scriptsize{}Encoder 4} & {\scriptsize{}Bottleneck} & {\scriptsize{}Decoder 1} & {\scriptsize{}Decoder 2} & {\scriptsize{}Output}\tabularnewline
		\midrule\midrule
		{\scriptsize{}Operation} & {\scriptsize{}Input} & {\scriptsize{}Conv+CBN} & {\scriptsize{}Conv+CBN} & {\scriptsize{}Conv+CBN} & {\scriptsize{}Conv+CBN} & {\scriptsize{}Conv} & {\scriptsize{}Upsample+Concat} & {\scriptsize{}Upsample+Concat} & {\scriptsize{}Conv}\tabularnewline
		\midrule
		{\scriptsize{}Channels} & {\scriptsize{}$1$} & {\scriptsize{}$1{\to}32{\to}128$} & {\scriptsize{}$128{\to}128$} & {\scriptsize{}$128{\to}256{\to}512$} & {\scriptsize{}$512{\to}1024$} & {\scriptsize{}$1024$} & {\scriptsize{}$2048{\to}1024{\to}512$} & {\scriptsize{}$1024{\to}512{\to}256$} & {\scriptsize{}$256{\to}1$}\tabularnewline
		\midrule
		{\scriptsize{}Size} & {\scriptsize{}$128{\times}64$} & {\scriptsize{}$128{\times}64$} & {\scriptsize{}$64{\times}32$} & {\scriptsize{}$32{\times}16$} & {\scriptsize{}$16{\times}8$} & {\scriptsize{}$8{\times}4$} & {\scriptsize{}$16{\times}8$} & {\scriptsize{}$64{\times}32$} & {\scriptsize{}$128{\times}64$}\tabularnewline
		\bottomrule
	\end{tabular}{\footnotesize\par}
\end{table*}

\begin{table*}[h]
	\caption{\label{table_complexity}Computational complexity of the proposed two-stage pipeline. Parameter counts are obtained directly from the released PyTorch models, and FLOPs are reported as $2\times$ the number of multiply--accumulate operations (i.e.\ $\text{FLOPs}=2\times\text{MACs}$) for a single $128\times63$ log-mel segment. Latencies are means over 1000 TensorRT FP16 inference calls on the target NVIDIA Jetson Orin Nano Super (MAXN Super power mode) with the models pre-loaded in GPU memory.}
	\centering{}{\footnotesize{}}%
	\begin{tabular}{lccc}
		\toprule
		\textbf{Stage} & \textbf{Parameters} & \textbf{FLOPs (per segment)} & \textbf{Latency (Orin Nano Super)}\tabularnewline
		\midrule\midrule
		Log-mel extraction (librosa) & --- & --- & $\approx2$~ms\tabularnewline
		N2N-Inspired Denoising Network & $\approx\ParamsDenoise$ & $\approx\FlopsDenoise$ FLOPs & $\approx15$~ms\tabularnewline
		Maneuver-CCAE & $\approx\ParamsCCAE$ & $\approx\FlopsCCAE$ FLOPs & $\approx2$~ms\tabularnewline
		Anomaly score + threshold & --- & --- & $<1$~ms\tabularnewline
		\textbf{Total} & $\boldsymbol{\approx}$\textbf{\ParamsTotal} & $\boldsymbol{\approx}$\textbf{\FlopsTotal} \textbf{FLOPs} & \textbf{$\approx20$~ms}\tabularnewline
		\bottomrule
	\end{tabular}{\footnotesize\par}
\end{table*}

\begin{table*}[h]
	\caption{\label{table3-1-1}Maneuver-Conditioned Convolutional Autoencoder architecture summary.}
	\centering{}{\footnotesize{}}%
	\begin{tabular}{cc>{\centering}m{1.5cm}>{\centering}m{3cm}>{\centering}m{2cm}>{\centering}m{2cm}>{\centering}m{1.5cm}>{\centering}m{1.5cm}}
		\toprule
		{\scriptsize{}Stage} & {\scriptsize{}Input} & {\scriptsize{}Embedding} & {\scriptsize{}Cond.\ Input} & {\scriptsize{}Encoder} & {\scriptsize{}Latent} & {\scriptsize{}Decoder} & {\scriptsize{}Output}\tabularnewline
		\midrule\midrule
		{\scriptsize{}Operation} & {\scriptsize{}Spectrogram} & {\scriptsize{}Conv layers} & {\scriptsize{}Concat with condition maps} & {\scriptsize{}Conv layers} & {\scriptsize{}Fully Connected} & {\scriptsize{}Conv layers} & {\scriptsize{}Conv}\tabularnewline
		\midrule
		{\scriptsize{}Channels} & {\scriptsize{}$1$} & {\scriptsize{}$16{\to}64$} & {\scriptsize{}$76$} & {\scriptsize{}$128{\to}256{\to}64{\to}16$} & {\scriptsize{}$16$} & {\scriptsize{}$16{\to}64{\to}256{\to}64{\to}16{\to}4$} & {\scriptsize{}$4{\to}1$}\tabularnewline
		\midrule
		{\scriptsize{}Size} & {\scriptsize{}$128{\times}64$} & {\scriptsize{}$64{\times}32$} & {\scriptsize{}$64{\times}32$} & {\scriptsize{}$4{\times}2$} & {\scriptsize{}$1{\times}1$} & {\scriptsize{}$128{\times}64$} & {\scriptsize{}$128{\times}64$}\tabularnewline
		\bottomrule
	\end{tabular}{\footnotesize\par}
\end{table*}

\subsection{Maneuver-Conditioned Convolutional Autoencoder}

Following the denoising stage, the refined log-mel spectrogram $\bar{X}_{i}\in\mathbb{R}^{1\times F\times T}$ is passed to a maneuver-CCAE that models the nominal acoustic behavior of the drone under different operating conditions. The conditional convolutional autoencoder uses a shallow, parameter-efficient architecture with an initial embedding stage, a conditioned convolutional encoder that compresses the input to a low-dimensional latent code, and a conditioned convolutional decoder, with CBN applied in the deeper encoder and decoder blocks. The architecture is summarized in Table~\ref{table3-1-1}.

\subsubsection{Conditional Input Embedding}

The denoised spectrogram $\bar{X}_{i}$ is mapped to an intermediate feature representation $E(\bar{X}_{i})$. Maneuver embeddings are reshaped into spatial maps and concatenated along the channel dimension, providing explicit maneuver information at every time-frequency location.

\subsubsection{Conditional Normalization}

In addition to input-level conditioning, maneuver information is injected into the encoder and decoder through CBN:
\begin{equation}
	\text{CBN}(Z|z_{i})=\gamma(z_{i})\hat{Z}+\beta(z_{i}).
\end{equation}

\subsubsection{Reconstruction and Anomaly Score}

The encoder maps the conditioned input to a latent representation, which the decoder maps back to the reconstruction $\hat{X}_{i}=g_{\phi}(\bar{X}_{i},z_{i})$. During inference, a test segment is first denoised by Stage~1 to give $\bar{X}^{*}$; with the associated maneuver embedding $z^{*}$, the anomaly score is the reconstruction error of the Stage~2 input:
\begin{equation}
	S^{*}=\frac{1}{FT}\sum_{f=1}^{F}\sum_{t=1}^{T}
	\!\left(\bar{X}^{*}(f,t)-\hat{X}^{*}(f,t)\right)^{2},
	\label{eq:anomaly_score}
\end{equation}
where $\hat{X}^{*}=g_{\phi}(\bar{X}^{*},z^{*})$ is the reconstruction of the denoised test input under its associated maneuver condition. The score is therefore computed between the test input presented to Stage~2 and its own reconstruction, so that no quantity from training enters the test-time score.

\subsection{Implementation Summary}

\paragraph*{Stage~1: Noise2Noise-Inspired Denoising}
\begin{enumerate}
	\item Convert raw audio to noisy log-mel spectrogram $\tilde{X}_{i}$ and collect maneuver labels $c_{i}^{\text{type}}$, $c_{i}^{\text{dir}}$.
	
	\item Project one-hot labels to 16-dimensional embedding $z_{i}$ ($d_{e}=8$).
	
	\item Extract features with CBN: $y_{b,c,h,w}=\gamma_{c}(z_{i})\hat{h}_{b,c,h,w}+\beta_{c}(z_{i})$.
	
	\item Optimize with hybrid loss $\mathcal{L}$ (Eq.~\eqref{eq:objective}, $\alpha=0.5$) against surrogate target $X_{i}^{\text{ref}}$.
	
	\item Output denoised spectrogram $\bar{X}_{i}$.
\end{enumerate}

\paragraph*{Stage~2: Maneuver-CCAE}
\begin{enumerate}
	\item Concatenate $E(\bar{X}_{i})$ with reshaped maneuver maps.
	
	\item Encode to latent bottleneck with CBN at every layer.
	
	\item Decode: $\hat{X}_{i}=g_{\phi}(\bar{X}_{i},z_{i})$.
	
	\item Compute anomaly score $S^{*}$.
	
	\item Classify using threshold $\tau$ (95th percentile of nominal validation reconstruction error):
	\begin{equation}
		\tau=\operatorname{Percentile}_{95}\!\left(\{S_i : X_i \in \mathcal{D}_{\text{val}}^{\text{nom}}\}\right),
	\end{equation}
	\begin{equation}
		\begin{cases}
			S^{*}>\tau & \Rightarrow \text{Faulty}, \\
			S^{*}\leq\tau & \Rightarrow \text{Normal}.
		\end{cases}
	\end{equation}
\end{enumerate}
\begin{figure*}[h]
	\includegraphics[width=\textwidth]{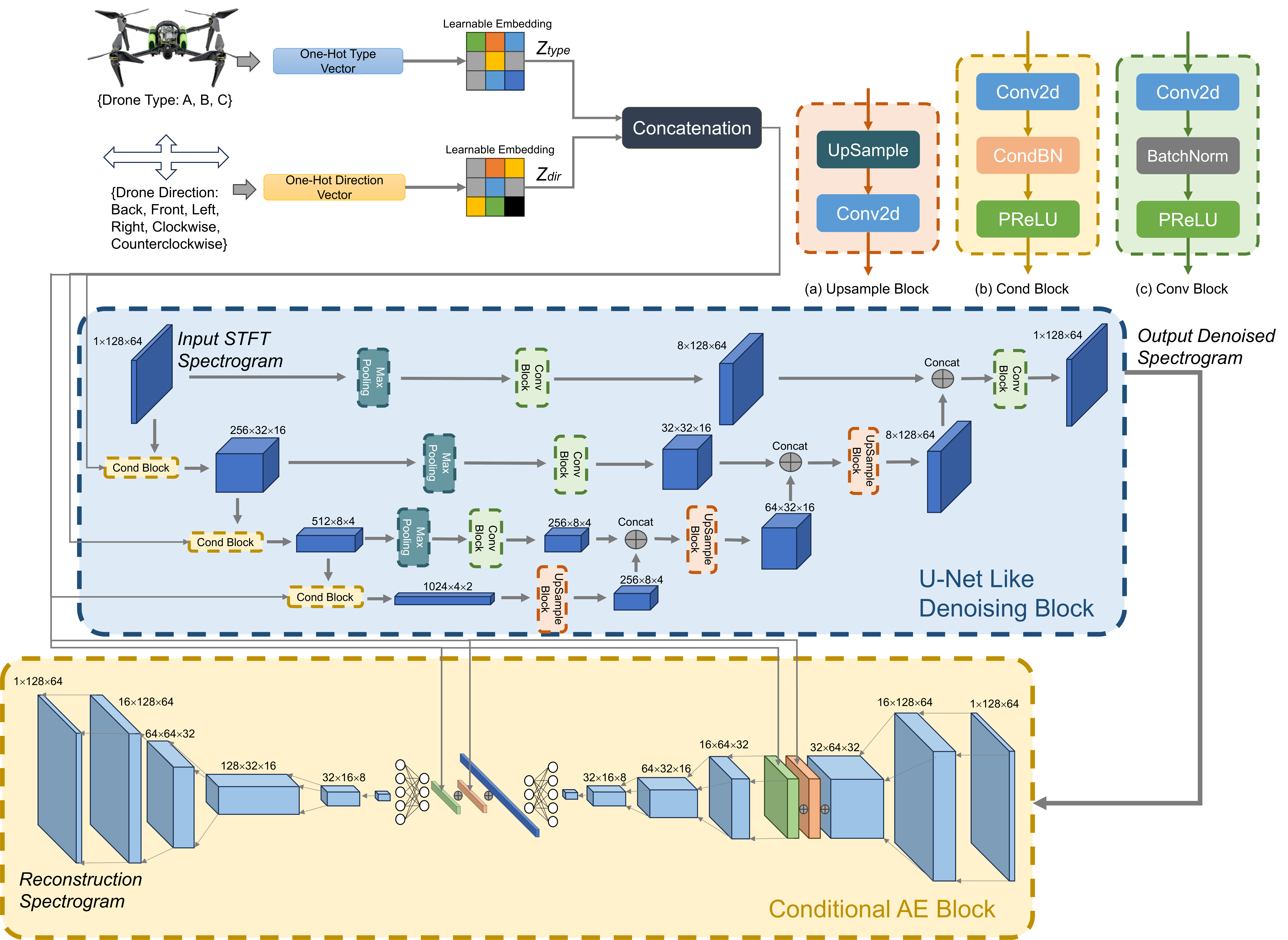}
	\caption{\label{fig:Proposed-two-stage-acoustic}Proposed two-stage acoustic fault detection framework. Stage~1: encoder-decoder denoising network. Stage~2: maneuver-conditioned convolutional autoencoder. The reconstruction error serves as the anomaly score.}
\end{figure*}

\section{Results and Discussion\label{sec: Results}}

\subsection{Experimental Setup}
All experiments use the dataset described in Section~\ref{sec:problem_formulation}. The full experimental protocol is presented to allow unambiguous interpretation of all subsequent metrics and figures.
\paragraph*{Dataset Composition} Three quadrotor platforms are used: Type~A (Holy Stone HS720), Type~B (MJX Bugs~12~EIS), and Type~C (ZLRC SG906 Pro2). Recordings cover six maneuver directions (forward, backward, left, right, clockwise, counterclockwise) and two health conditions (normal and faulty). Fault conditions include propeller blade partial cuts and motor cap mechanical deformation. The recordings were originally captured at 48~kHz and downsampled to 16~kHz, then segmented into non-overlapping 2-second windows ($N=32{,}000$ samples per segment). The dataset composition, following the ICSV31 AI Challenge partition, is summarized in Table~\ref{table_dataset}: a training split of 5{,}400 normal segments (1{,}800 per drone type), an evaluation split of 1{,}080 segments containing both normal and anomalous data spanning six fault types (360 segments per drone type, that is, 180 segments per fault type), and a test split of 1{,}440 anomalous segments spanning eight fault types (480 segments per drone type, that is, 180 segments per fault type).
\paragraph*{Data Splitting} This work adopts the predefined dataset partition rather than constructing a custom split. The model is trained exclusively on the training split, which contains normal segments only, consistent with the unsupervised (self-supervised) setting. For threshold selection, a nominal validation subset is held out from the training split; no anomalous data are used at any point during training or threshold selection. All reported detection metrics (AUC, precision, recall, F1-score, FPR, FNR, and the per-drone-type and per-maneuver breakdowns) are computed on the evaluation split, which contains both normal and anomalous segments and is disjoint from the training data. The test split of the challenge partition contains anomalous segments only and therefore admits neither a false-positive rate nor an AUC; it is listed in Table~\ref{table_dataset} for completeness but is not used for any reported metric. This partition follows the dataset's own train/eval separation and therefore avoids any leakage of evaluation data into training.
\paragraph*{Threshold Selection} The anomaly threshold $\tau$ is selected using only the held-out nominal validation subset of the training split ($\mathcal{D}_{\text{val}}^{\text{nom}}$), without any access to faulty samples:
\begin{equation}
	\tau=\operatorname{Percentile}_{95}\!\left(\{S_{i}:X_{i}\in
	\mathcal{D}_{\text{val}}^{\text{nom}}\}\right).
\end{equation}
Concretely, the reconstruction error $S_{i}$ of Eq.~\eqref{eq:anomaly_score} is evaluated for every segment of $\mathcal{D}_{\text{val}}^{\text{nom}}$, the resulting values are sorted in ascending order as $S_{(1)}\leq S_{(2)}\leq\cdots\leq S_{(N_{v})}$, the index $k=\lceil0.95N_{v}\rceil$ is formed, and the threshold is set to $\tau=S_{(k)}$. No faulty segment is inspected and no label is read at any point in this procedure.
The 95th percentile is a conservative operating choice rather than a tuned parameter: by construction, 95\% of nominal segments fall below $\tau$ and the remaining 5\% are expected to raise false alarms. For UAV health monitoring a missed fault is more costly than a false alarm, which motivates a tolerance on the permissive side. A sensitivity analysis over the 90th--99th percentile range confirms that detection performance is stable across this interval.
The complete pipeline is therefore unsupervised in three distinct senses: no faulty segment is used for training, no faulty label is used for threshold selection, and no supervised optimization of the threshold is performed. The threshold follows solely from the statistics of the nominal validation error distribution.

\begin{table*}[h]
	\centering{}\caption{\label{table_dataset} Dataset composition by drone type, following the ICSV31 AI Challenge partition. Each file is a non-overlapping 2-second segment (recorded at 48~kHz, downsampled to 16~kHz). The \emph{train} split contains normal segments only; the \emph{eval} split contains both normal and anomalous segments spanning six fault types (three propeller, three motor); the \emph{test} split contains anomalous segments spanning eight fault types (two additional) and, containing no normal segments, is not used for the reported detection metrics. All platforms are recorded across the six maneuver directions.}
	\begin{tabular}{lccccc}
		\toprule
		\textbf{Platform} & \textbf{Directions} & \textbf{Train (normal)} & \textbf{Eval (normal+faulty)} & \textbf{Test (faulty)} & \textbf{Total}\tabularnewline
		\midrule\midrule
		Type~A (Holy Stone HS720) & 6 & 1{,}800 & 360 & 480 & 2{,}640\tabularnewline
		Type~B (MJX Bugs~12~EIS) & 6 & 1{,}800 & 360 & 480 & 2{,}640\tabularnewline
		Type~C (ZLRC SG906 Pro2) & 6 & 1{,}800 & 360 & 480 & 2{,}640\tabularnewline
		\midrule
		\textbf{Total} & 6 & \textbf{5{,}400} & \textbf{1{,}080} & \textbf{1{,}440} & \textbf{7{,}920}\tabularnewline
		\bottomrule
	\end{tabular}
\end{table*}

\begin{table*}[t]
	\centering
	\caption{%
		Left: Extended performance metrics for the proposed full model (overall);
		Right: Ablation study contribution of each pipeline stage. The ablation
		configurations correspond one-to-one with rows of Table~\ref{table_baseline}:
		\textit{AE only} is \textit{AE-1 (Orig)}, \textit{Denoising + AE} is
		\textit{AE-1 (Denoised)}, and \textit{AE + Conditioning} is
		\textit{CCAE (Ours, Orig)}.
	}
	\label{table_metrics_ablation}
	
	\renewcommand{\arraystretch}{1.2}
	\setlength{\tabcolsep}{4pt}
	
	\begin{adjustbox}{max width=\textwidth}
		\begin{tabular}{
				@{}
				lccccccc
				@{\hspace{0.8em}\vrule\hspace{0.8em}}
				lcccc
				@{}
			}
			\toprule
			\multicolumn{8}{c}{\textbf{(a) Extended Performance Metrics}}
			&
			\multicolumn{5}{c}{\textbf{(b) Ablation Study}}
			\\
			\midrule
			
			\textbf{Model}
			& \textbf{AUC}
			& \textbf{Acc.}
			& \textbf{Prec.}
			& \textbf{Recall}
			& \textbf{F1}
			& \textbf{FPR}
			& \textbf{FNR}
			&
			\textbf{Configuration}
			& \textbf{AUC}
			& \textbf{Prec.}
			& \textbf{Recall}
			& \textbf{F1}
			\\
			\midrule
			
			CCAE (Orig)
			& \AUCcondOnly
			& 0.623
			& \PREccondOnly
			& \RECcondOnly
			& \FoneCondOnly
			& 0.046
			& 0.707
			&
			AE only (no cond., no denoising)
			& \AUCaeOnly
			& \PRECaeOnly
			& \RECaeOnly
			& \FoneAeOnly
			\\
			
			\textbf{Full model (Denoised)}
			& \textbf{\AUCfull}
			& \textbf{0.806}
			& \textbf{0.754}
			& \textbf{0.907}
			& \textbf{0.824}
			& \textbf{0.296}
			& \textbf{0.093}
			&
			AE + Conditioning (i.e.\ CCAE, Orig)
			& \AUCcondOnly
			& \PREccondOnly
			& \RECcondOnly
			& \FoneCondOnly
			\\
			
			& & & & & & & &
			Denoising + AE (no cond.)
			& \AUCdenoiseAE
			& \PRECdenoiseAE
			& \RECdenoiseAE
			& \FoneDenoiseAE
			\\
			
			& & & & & & & &
			\textbf{Full model (N2N + CCAE)}
			& \textbf{\AUCfull}
			& \textbf{0.754}
			& \textbf{0.907}
			& \textbf{0.824}
			\\
			
			\bottomrule
		\end{tabular}
	\end{adjustbox}
\end{table*}

\begin{table}[h]
	\centering
	\caption{\label{table_confusion}Row-normalized confusion matrix of the proposed full model on the evaluation split, obtained from the reported recall (TPR~$=0.907$) and false-positive rate (FPR~$=0.296$).}
	\renewcommand{\arraystretch}{1.2}
	\begin{tabular}{llcc}
		\toprule
		& & \multicolumn{2}{c}{\textbf{Predicted}}\tabularnewline
		\cmidrule(l){3-4}
		& & \textbf{Normal} & \textbf{Faulty}\tabularnewline
		\midrule\midrule
		\multirow{2}{*}{\textbf{Actual}} & \textbf{Normal} & 0.704 & 0.296\tabularnewline
		& \textbf{Faulty} & 0.093 & 0.907\tabularnewline
		\bottomrule
	\end{tabular}
\end{table}

\begin{table*}[t]
	\centering
	\caption{\label{table_baseline}%
		Comparison of unsupervised anomaly detection methods.
		\textit{Orig} = original log-mel spectrograms;
		\textit{Denoised} = spectrograms processed by the proposed N2N-inspired stage.
		The suffix \textit{-1} denotes a single global model fitted to all operating
		conditions, and \textit{-18} denotes a bank of 18 condition-specific models
		(3 drone types $\times$ 6 maneuver directions). All methods share the same
		train/eval partition, log-mel features ($F\times T=128\times 63$),
		95th-percentile threshold, and evaluation protocol.}
	\renewcommand{\arraystretch}{1.15}
	\begin{tabular*}{\textwidth}{@{\extracolsep{\fill}}llccccc@{}}
		\toprule
		\textbf{Category} & \textbf{Method} & \textbf{AUC} & \textbf{Accuracy} & \textbf{Precision} & \textbf{Recall} & \textbf{F1}\\
		\midrule
		\multirow{6}{*}{\shortstack[l]{Traditional\\ML\\(single)}}
		& IF-1 (Orig)          & 0.607 & 0.562 & 0.868 & 0.146 & 0.250 \\
		& IF-1 (Denoised)      & 0.776 & 0.653 & 0.807 & 0.402 & 0.536 \\
		& GMM-1 (Orig)         & 0.605 & 0.578 & 0.882 & 0.180 & 0.298 \\
		& GMM-1 (Denoised)     & 0.782 & 0.651 & 0.796 & 0.406 & 0.537 \\
		& OC-SVM-1 (Orig)      & 0.605 & 0.586 & 0.872 & 0.202 & 0.328 \\
		& OC-SVM-1 (Denoised)  & 0.782 & 0.673 & 0.781 & 0.481 & 0.596 \\
		\midrule
		\multirow{6}{*}{\shortstack[l]{Traditional\\ML\\(conditioned)}}
		& IF-18 (Orig)         & 0.664 & 0.581 & 0.866 & 0.191 & 0.313 \\
		& IF-18 (Denoised)     & 0.856 & 0.716 & 0.828 & 0.544 & 0.657 \\
		& GMM-18 (Orig)        & 0.688 & 0.622 & 0.840 & 0.301 & 0.444 \\
		& GMM-18 (Denoised)    & 0.865 & 0.768 & 0.826 & 0.678 & 0.745 \\
		& OC-SVM-18 (Orig)     & 0.666 & 0.620 & 0.696 & 0.428 & 0.530 \\
		& OC-SVM-18 (Denoised) & 0.832 & 0.760 & 0.739 & 0.804 & 0.770 \\
		\midrule
		\multirow{10}{*}{\shortstack[l]{Reconstruction-\\based AE}}
		& AE-1 (Orig)          & \AUCaeOnly & 0.596 & \PRECaeOnly & \RECaeOnly & \FoneAeOnly \\
		& AE-1 (Denoised)      & \AUCdenoiseAE & 0.587 & \PRECdenoiseAE & \RECdenoiseAE & \FoneDenoiseAE \\
		& DAE-1 (Orig)         & 0.623 & 0.597 & 0.766 & 0.280 & 0.410 \\
		& DAE-1 (Denoised)     & 0.776 & 0.678 & 0.798 & 0.476 & 0.596 \\
		& VAE-1 (Orig)         & 0.602 & 0.581 & 0.759 & 0.239 & 0.363 \\
		& VAE-1 (Denoised)     & 0.648 & 0.563 & 0.770 & 0.180 & 0.291 \\
		& MaskedAE-1 (Orig)    & 0.666 & 0.630 & 0.789 & 0.354 & 0.488 \\
		& MaskedAE-1 (Denoised)& 0.780 & 0.669 & 0.808 & 0.444 & 0.573 \\
		& MemAE-1 (Orig)       & 0.624 & 0.611 & 0.800 & 0.296 & 0.432 \\
		& MemAE-1 (Denoised)   & 0.794 & 0.681 & 0.838 & 0.450 & 0.586 \\
		\midrule
		\multirow{6}{*}{\shortstack[l]{State-of-the-art\\frameworks}}
		& SimpleNet (Orig)     & 0.591 & 0.588 & 0.556 & 0.873 & 0.679 \\
		& SimpleNet (Denoised) & 0.611 & 0.572 & 0.543 & 0.917 & 0.682 \\
		& Dinomaly (Orig)      & 0.606 & 0.507 & 0.504 & 0.993 & 0.669 \\
		& Dinomaly (Denoised)  & 0.790 & 0.697 & 0.649 & 0.861 & 0.740 \\
		& AADCL (Orig)         & 0.571 & 0.522 & 0.609 & 0.124 & 0.206 \\
		& AADCL (Denoised)     & 0.706 & 0.613 & 0.742 & 0.346 & 0.472 \\
		\midrule
		\multirow{2}{*}{\shortstack[l]{\textbf{Proposed}}}
		& CCAE (Ours, Orig)  & \AUCcondOnly & 0.623 & \PREccondOnly & \RECcondOnly & \FoneCondOnly \\
		& \textbf{CCAE (Ours, Denoised)} & \textbf{\AUCfull} & \textbf{0.806} & \textbf{0.754} & \textbf{0.907} & \textbf{0.824} \\
		\bottomrule
	\end{tabular*}
\end{table*}

\subsection{Detection Performance}
\begin{figure*}[h]
	\subfloat[ROC curve.]{\includegraphics[width=9cm,height=6cm]{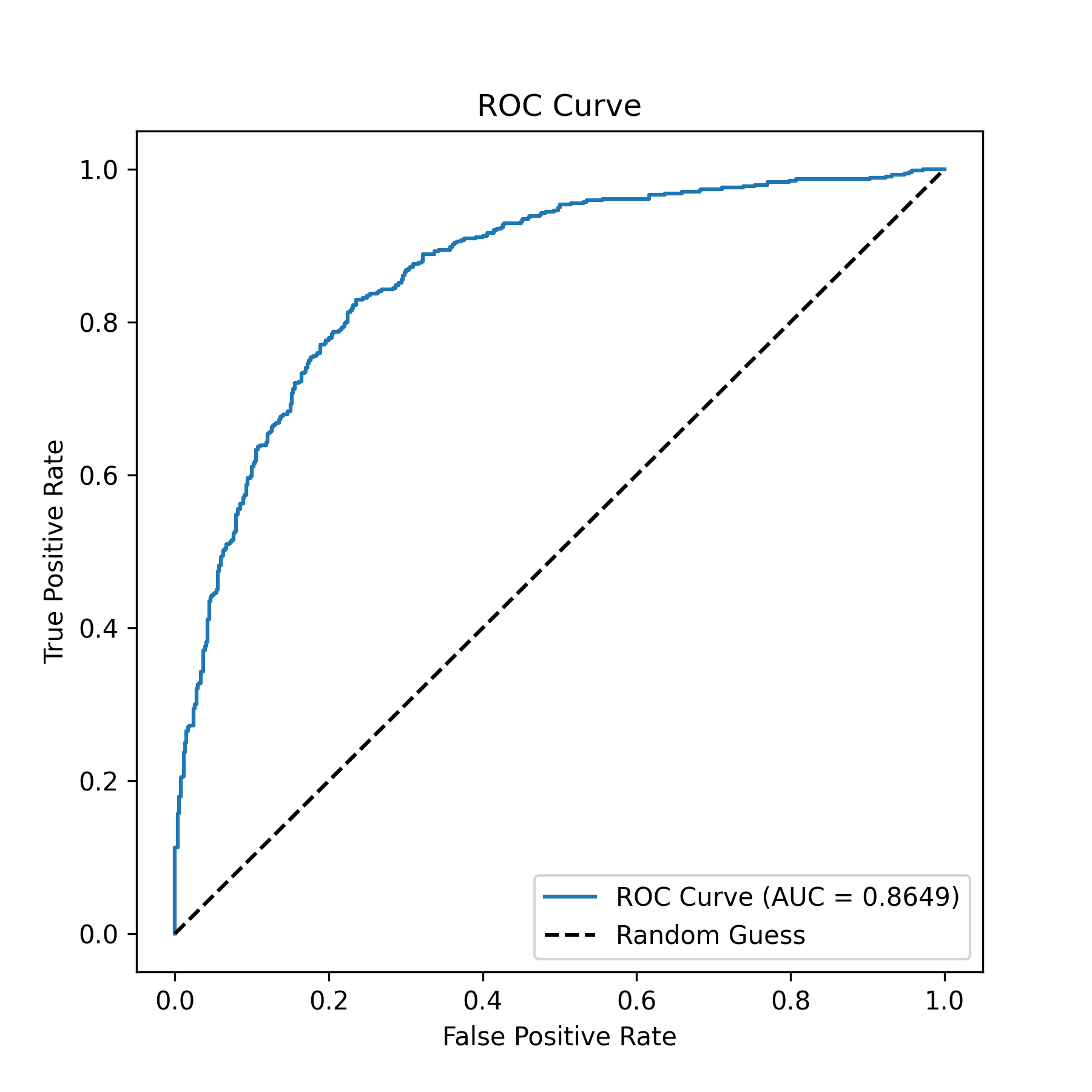}}
	\hfill{}
	\subfloat[Training convergence.]{\includegraphics[width=9cm,height=9cm,keepaspectratio]{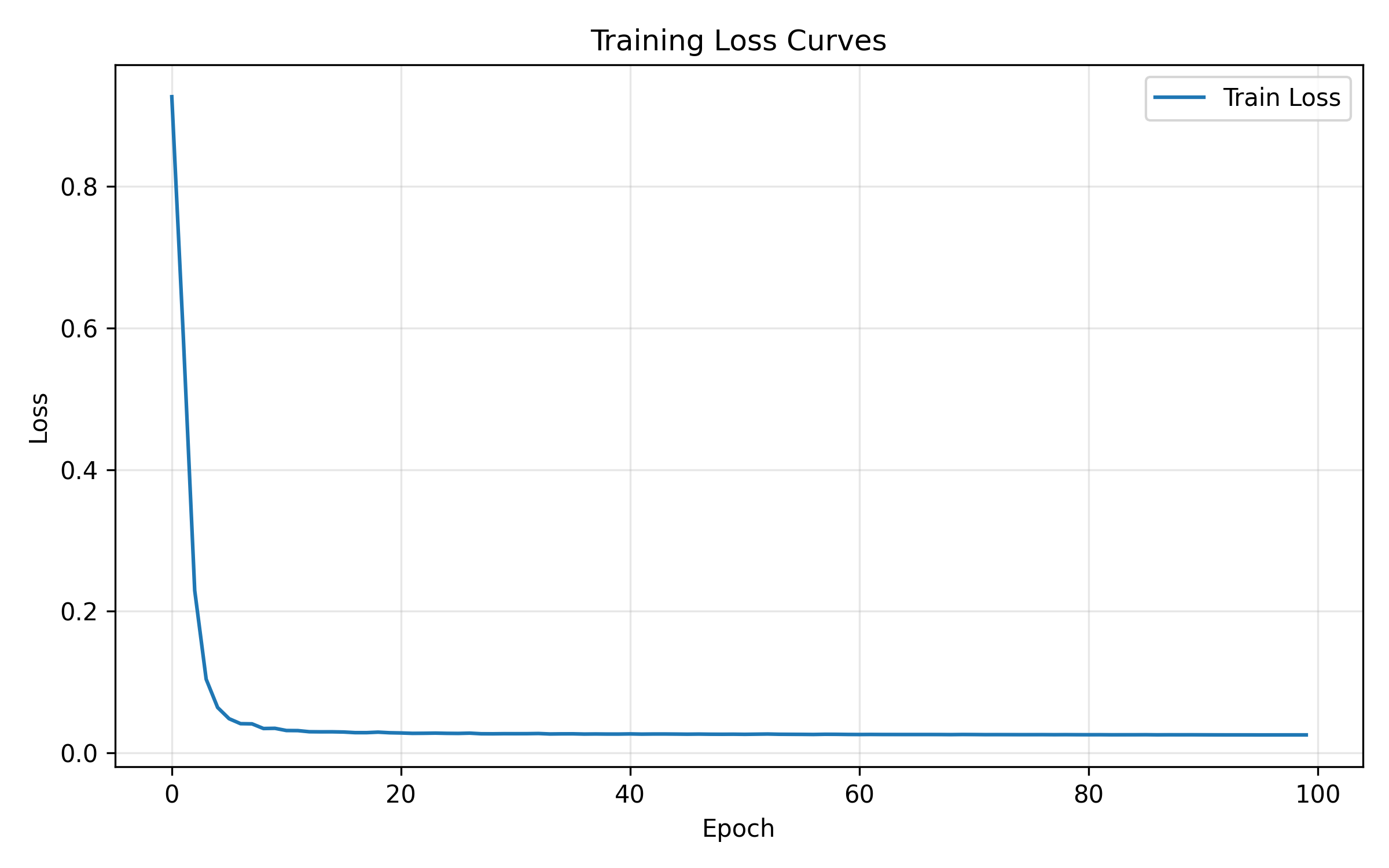}}
	\hfill{}
	\subfloat[3D t-SNE of latent representations.]{\includegraphics[width=9cm,height=9cm,keepaspectratio]{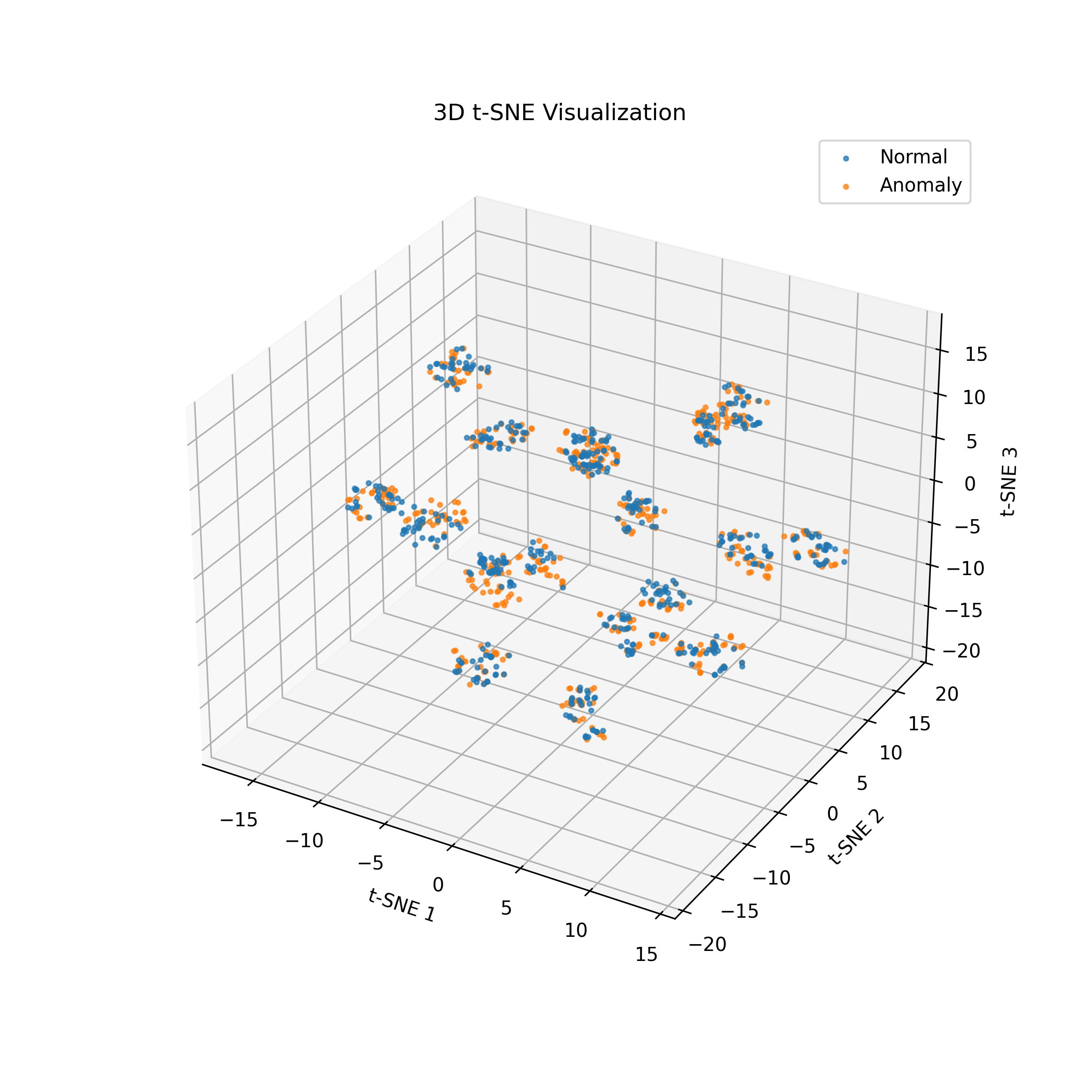}}
	\hfill{}
	\subfloat[Reconstruction error distributions (log scale).]{\includegraphics[width=9cm,height=9cm,keepaspectratio]{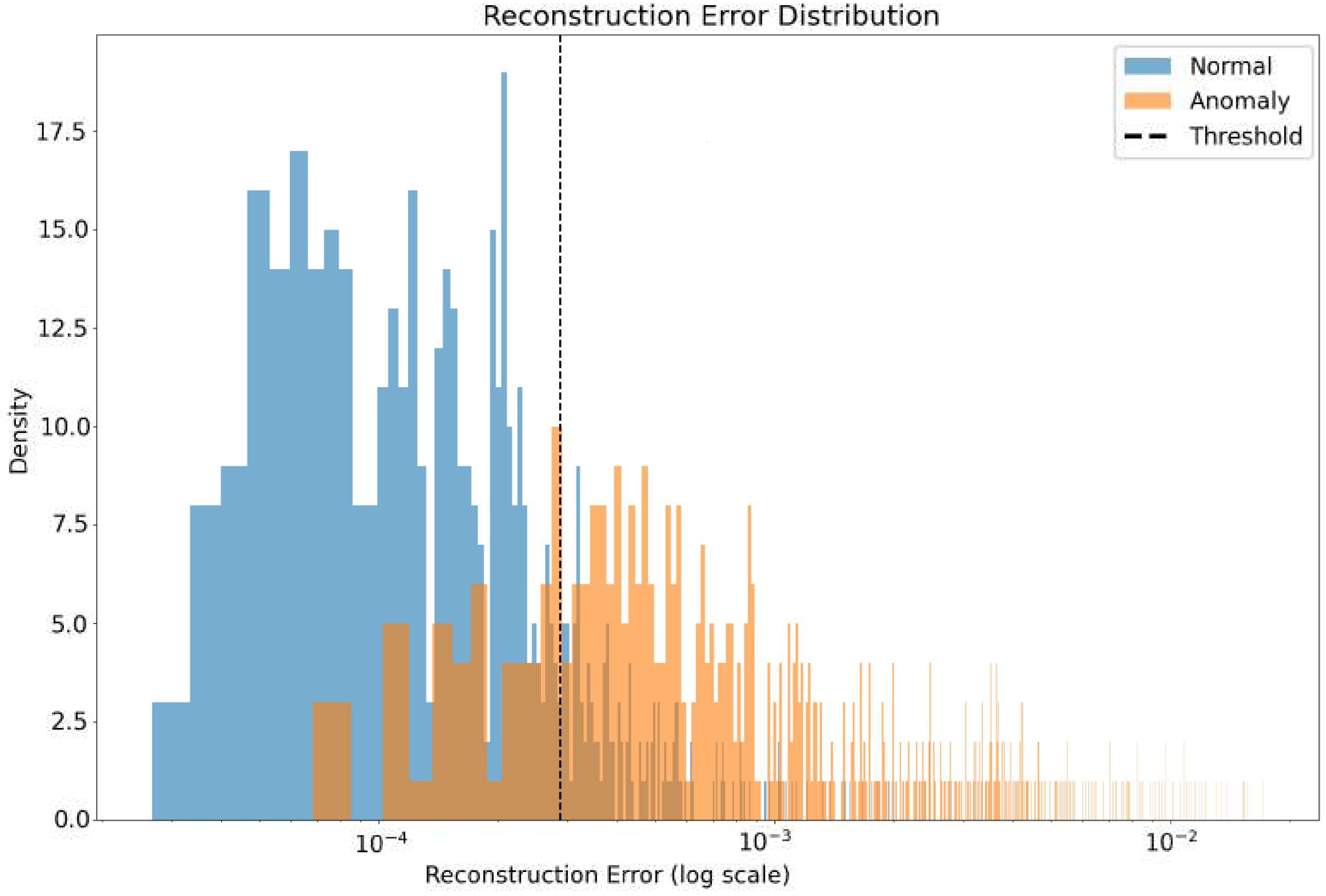}}
	\caption{\label{fig:results}Performance evaluation. $(a)$~ROC curve (AUC~$=\AUCfull$). $(b)$~Training loss convergence. $(c)$~t-SNE latent clustering. $(d)$~Reconstruction error distributions for nominal and faulty observations.}
\end{figure*}

The proposed maneuver-conditioned acoustic fault detection framework performance is evaluated using reconstruction-error separability, latent representation structure, and convergence characteristics. Since the model is trained using only nominal acoustic data, the evaluation emphasizes the ability of the learned representation to distinguish normal maneuver-induced variability from fault-induced deviations. Fig.~\ref{fig:results}(d) shows a clear statistical separation between nominal and faulty reconstruction error distributions, spanning multiple orders of magnitude. Fig.~\ref{fig:results}(a) shows the ROC curve with AUC~$=\AUCfull$ for the full model (N2N-inspired denoising + CCAE), indicating strong separability. The visualization in Fig.~\ref{fig:results}(c) reveals that nominal acoustic samples form compact clusters corresponding to maneuver-dependent operating conditions, while faulty samples appear as outliers. This confirms that the conditioning mechanism successfully separates structured nominal variability from fault-induced deviations. Fig.~\ref{fig:results}(b) illustrates stable training loss convergence.
\paragraph*{Extended Evaluation Metrics} In addition to AUC, Table~\ref{table_metrics_ablation} reports precision, recall (True Positive Rate, TPR), F1-score, False Positive Rate (FPR), and False Negative Rate (FNR) for the proposed full model. The most significant observation is the reduction of the False Negative Rate from 0.707 (CCAE without denoising) to 0.093 (full model), corresponding to an increase in recall from 0.293 to 0.907. Although the full model exhibits a higher FPR (0.296 vs.~0.046), the resulting operating point is substantially more suitable for UAV fault monitoring, where missed fault detections are generally more critical than false alarms. Results are broken down by drone type: Type~C achieves the highest recall (0.972) and the lowest FNR (0.028), indicating particularly strong fault separability, while all platforms achieve AUC values above 0.86. Across all six maneuver directions, AUC values remain within 0.831--0.914 and F1-scores within 0.782--0.856. The clockwise maneuver is the most challenging condition due to higher aerodynamic noise masking fault signatures. Both injected fault categories, the propeller blade cuts and the motor-cap deformations, are detected across all three platforms, so the detector responds to generic fault-induced spectral deviations rather than to a single fault signature. The corresponding row-normalized confusion matrix of the full model is reported in Table~\ref{table_confusion}.

\subsection{Comparison with Unsupervised Baselines}

Since the proposed framework is inherently unsupervised and requires only healthy nominal data for training, comparisons are performed exclusively against unsupervised methods operating under the same assumptions. All baselines are evaluated under identical experimental conditions (same log-mel features $F\times T=128\times 63$, same train/eval partition, same 95th-percentile threshold). Table~\ref{table_baseline} reports the full results. Several important observations can be drawn.

First, the proposed denoising module consistently improves almost all anomaly detectors. Replacing original spectrograms with denoised spectrograms leads to substantial AUC improvements across all paradigms: IF-1 improves from 0.607 to 0.776, GMM-1 from 0.605 to 0.782, MemAE-1 from 0.624 to 0.794, and AADCL from 0.571 to 0.706, demonstrating that the denoising benefit generalizes across fundamentally different detection paradigms. The gain is not uniform across every metric: for AE-1 and VAE-1 the F1-score falls slightly even though the AUC rises, because the fixed 95th-percentile threshold no longer sits at the same point of the score distribution. Second, explicitly modeling operating-condition variability is more important than increasing detector complexity. Simply training condition-specific models (18 models for 3 drone types $\times$ 6 maneuvers) consistently outperforms single global models. The denoised GMM-18 (AUC~$=0.865$) surpasses every single global model in the table, including the strongest deep-learning baseline (MemAE-1, AUC~$=0.794$) and the strongest recent framework (Dinomaly, AUC~$=0.790$), strongly suggesting that maneuver-dependent variability is the dominant challenge in this dataset. Third, conventional reconstruction-based autoencoders (AE, DAE, VAE, MaskedAE, MemAE) struggle with the highly heterogeneous nominal distribution because they assume a single latent representation across all 18 operating regimes. Fourth, SimpleNet and Dinomaly designed for RGB-image anomaly detection do not transfer effectively to UAV acoustic spectrograms. Fifth, AADCL's contrastive feature space is dominated by operating-condition variability, reducing its anomaly discrimination. Finally, the proposed CCAE with denoising achieves the highest overall AUC ($\AUCfull$) and F1-score (0.824), surpassing the strongest traditional baseline (GMM-18: AUC~$=0.865$, F1~$=0.745$) while using a single conditioned model in place of 18 condition-specific ones, validating the proposed maneuver-aware two-stage architecture.

\subsection{Embedded Real-Time Implementation}
To demonstrate practical viability, the trained model was integrated into a ROS2-based onboard perception pipeline and executed on an NVIDIA Jetson Orin Nano embedded computer.
\paragraph*{Hardware} NVIDIA Jetson Orin Nano Super Developer Kit; 6-core Arm Cortex-A78AE (up to 1.7~GHz); 1024-core NVIDIA Ampere GPU with 32 third-generation Tensor Cores; 8~GB LPDDR5 (102~GB/s); NVMe SSD via M.2; JetPack~6.1 (Ubuntu~22.04 LTS, 64-bit ARM); MAXN Super power mode; ROS2 Humble Hawksbill middleware.
\paragraph*{Software and Numerical Precision} Python~3.10; PyTorch~2.x for training. For deployment, the trained Stage~1 and Stage~2 models are exported to ONNX and compiled into TensorRT half-precision (FP16) inference engines that execute on the GPU Tensor Cores. No fixed-point quantization is applied in the current prototype; INT8 post-training quantization is reserved for future work, which could provide a further reduction in both latency and memory footprint.
\begin{itemize}
	\item \textit{Audio Acquisition. } R{\O}DE Wireless Go2 microphone; $f_{s}=16{,}000$~Hz (16-bit PCM); 2-second segments ($N=32{,}000$ samples); non-overlapping windows.
	\item \textit{ROS2 Node Architecture.  The pipeline comprises two \texttt{rclpy} (Python) nodes: (i)~the \textit{audio\_preprocessor} node, which subscribes to the \texttt{/audio/raw} topic, segments the incoming audio stream into 2-second windows, computes the log-mel spectrogram $X_{\text{log-mel}}\in\mathbb{R}^{F\times T}$, and publishes it to \texttt{/audio/logmel}; and (ii)~the \textit{fault\_detector} node, which subscribes to \texttt{/audio/logmel} and \texttt{/uav/maneuver}, runs the Noise2Noise-inspired denoising network and the CCAE sequentially as pre-built TensorRT FP16 engines on the GPU (with pinned host buffers and a persistent CUDA stream to minimize host-side overhead), computes the anomaly score $S^{*}$, compares it against the pre-loaded threshold $\tau$, and publishes the anomaly score and binary fault flag to \texttt{/fault/detection}. The inter-node communication via ROS2's DDS middleware adds negligible latency ($<0.1$~ms).}
\end{itemize}
The per-segment latency breakdown is reported in Table~\ref{table_complexity}. The total end-to-end latency of $\approx20$~ms is well below the 2-second audio window duration, confirming that the system operates in real time without buffer accumulation under continuous operation. These measurements were obtained by averaging over 1000 TensorRT FP16 inference calls on the Jetson Orin Nano Super, with the engines pre-loaded into GPU memory.

\section{Conclusion\label{sec: Conclusion}}
This paper presented a maneuver-conditioned acoustic fault detection framework for autonomous quadrotor drones that combines Noise2Noise-inspired denoising with conditional reconstruction learning under a strictly unsupervised paradigm. The proposed approach addresses two fundamental challenges in real-flight acoustic monitoring: stochastic environmental noise and structured variability induced by maneuver-dependent aerodynamic effects. Unlike conventional acoustic fault detection methods that assume stationary operating conditions or rely on extensive labeled fault datasets, the proposed framework explicitly incorporates maneuver context into both the denoising and reconstruction stages, enabling the model to distinguish between benign operational variability and genuine fault-induced deviations. The proposed architecture employs a two-stage unsupervised learning strategy. First, a Noise2Noise-inspired denoising network suppresses environmental disturbances using maneuver-consistent surrogate pairs, without requiring clean reference signals. Second, a maneuver-conditioned convolutional autoencoder models nominal acoustic behavior using only healthy-flight data, with both stages conditioning on drone type and flight direction via Conditional Batch Normalization. The anomaly threshold is selected entirely from a held-out nominal validation subset of the training split, ensuring complete unsupervised compliance. Experimental results demonstrate that the proposed maneuver-aware conditioning raises the AUC from $\AUCaeOnly$ (unconditioned baseline) to $\AUCfull$ (full model), a gain of $+0.323$ attributable to the combined effect of denoising and maneuver conditioning. A comparison against 11 unsupervised baselines spanning classical methods (IF, GMM, OC-SVM), reconstruction-based methods (AE, DAE, VAE, MaskedAE, MemAE), contrastive methods (AADCL), and recent state-of-the-art frameworks (SimpleNet, Dinomaly) confirms that the proposed full model attains the highest AUC and F1-score of all evaluated methods. The framework achieves an end-to-end inference latency of approximately 20~ms on an embedded NVIDIA Jetson Orin Nano Super with a TensorRT FP16 backend, demonstrating practical viability for real-time onboard UAV health monitoring.

\balance
\bibliographystyle{IEEEtran}
\bibliography{Bib_UAV_Fault}
		
	\end{document}